\documentclass[preprint,10pt,final]{elsarticle}
\usepackage{etex}

\usepackage{amssymb,amsbsy,exscale,amsmath,amsfonts,amscd,amsthm}
\usepackage[letterpaper]{geometry}
\usepackage[usenames,dvipsnames]{color}
\usepackage{bm}
\usepackage{soul}
\usepackage[ampersand]{easylist}
\usepackage{graphicx}
\graphicspath{{./images/}}
\usepackage{parskip}
\usepackage{longtable}
\usepackage{subcaption}
\usepackage[usenames,dvipsnames]{color}
\usepackage{multirow}
\usepackage{adjustbox}
\usepackage{algorithmicx}
\usepackage[ruled]{algorithm}
\usepackage{algpseudocode}
\biboptions{sort&compress}
\usepackage{pifont}
\usepackage{tikz}
\usetikzlibrary{mindmap,backgrounds}
\usepackage{epstopdf}
\usepackage[nottoc]{tocbibind}
\usepackage{stmaryrd}
\usepackage{mathrsfs}
\usepackage[format=hang,font=footnotesize]{caption}
\usepackage[format=hang,font=footnotesize]{subcaption}
\usepackage{epigraph}
\usepackage{array}
\usepackage{booktabs,makecell}
\usepackage{parskip}
\usepackage{mathtools}
\usepackage[mathscr]{euscript}
\usepackage{upgreek}
\usepackage[overload]{empheq}
\usepackage{xspace}
\usepackage{arydshln}
\usepackage{cmap}
\usepackage[T1]{fontenc}
\usepackage{esint}
\usepackage{dsfont}

\newcommand{\lb}{\left[}
\newcommand{\rb}{\right]}
\newcommand{\lp}{\left(}
\newcommand{\rp}{\right)}

\DeclareCaptionLabelFormat{andtable}{#1~#2 \& \tablename~\thetable}

\newcommand{\mbb}[1]{\mathbb{#1}}

\newcommand{\mcal}[1]{\mathcal{#1}}

\newcommand{\abs}[1]{\left\lvert{#1}\right\rvert}

\newcommand{\mG}{\bm G}

\newcommand{\refdom}{\mcal{D}_{\bf r}}
\newcommand{\phdom}{\mcal{D}}
\newcommand{\mcT}[1]{\mathcal{T}_h^{#1}}
\newcommand{\bx}{{\bf x}}

\newcommand{\bn}{{\bf n}}
\newcommand{\br}{{\bf r}}
\newcommand{\bt}{{\bf t}}

\newcommand{\refN}{{\bf n}_{\bf r}}
\newcommand{\refX}{{\bf x}_{\bf r}}
\newcommand{\refV}{\nabla_{\bf r}}

\newcommand{\bE}{{\bf E}}
\newcommand{\bH}{{\bf H}}
\newcommand{\bV}{{\bf V}}

\newcommand{\bv}{{\bf v}}

\newcommand{\bc}{{\bf c}}
\newcommand{\bu}{{\bf u}}
\newcommand{\be}{{\bf e}}

\newcommand{\mm}{$\upmu$m\xspace}
\newcommand{\upa}{\bm \uptheta}

\newcommand{\ns}{\dagger}
\newcommand{\rd}{\dagger}
\newcommand{\snap}{{\bm \zeta}^\ns}
\newcommand{\snapv}{{\bm \zeta}^\rd}
\newcommand{\pow}{\varsigma}
\newcommand{\fie}{\pi}

\newcommand{\AO}{\mbox{Al}_2\mbox{O}_3}

\makeatletter
\newcommand*\rel@kern[1]{\kern#1\dimexpr\macc@kerna}
\newcommand*\widebar[1]{%
  \begingroup
  \def\mathaccent##1##2{%
    \rel@kern{0.8}%
    \overline{\rel@kern{-0.8}\macc@nucleus\rel@kern{0.2}}%
    \rel@kern{-0.2}%
  }%
  \macc@depth\@ne
  \let\math@bgroup\@empty \let\math@egroup\macc@set@skewchar
  \mathsurround\z@ \frozen@everymath{\mathgroup\macc@group\relax}%
  \macc@set@skewchar\relax
  \let\mathaccentV\macc@nested@a
  \macc@nested@a\relax111{#1}%
  \endgroup
}
\makeatother

\newcommand{\Linux}{\texttt{Linux}\xspace}

\newcommand{\epAO}{\varepsilon_{\tiny\AO}}

 \def\etal{{\it et al.}\xspace}
\def\eg{{\it e.g.}\xspace}

\def\apo{{\it a posteriori}\xspace}
\def\apr{{\it a priori}\xspace}

\newcommand*{\largeCdot}{\raisebox{-0.25ex}{\scalebox{1.4}{$\cdot$}}}

\journal{Journal of Computational Physics}

\begin{document}

\begin{frontmatter}

\title{Computing parametrized solutions for plasmonic nanogap structures}

\author[mit]{F.~Vidal-Codina\corref{cor1}}
\ead{fvidal@mit.edu}
\author[mit]{N.~C.~Nguyen}
\ead{cuongng@mit.edu}
\author[mit]{J.~Peraire}
\ead{peraire@mit.edu}

\cortext[cor1]{Corresponding author}
 \address[mit]{Dept. of Aeronautics and Astronautics, Massachusetts Institute of Technology, Cambridge, MA 02139, USA}

 \begin{abstract}
The interaction of electromagnetic waves with metallic nanostructures generates resonant oscillations of the conduction-band electrons at the metal surface. These resonances can lead to large enhancements of the incident field and to the confinement of light to small regions, typically several orders of magnitude smaller than the incident wavelength. The accurate prediction of these resonances entails several challenges. Small geometric variations in the plasmonic structure may lead to large variations in the electromagnetic field responses. Furthermore, the material parameters that characterize the optical behavior of metals at the nanoscale need to be determined experimentally and are consequently subject to measurement errors. It then becomes essential that any predictive tool for the simulation and design of plasmonic structures accounts for fabrication tolerances and measurement uncertainties.

In this paper, we develop a reduced order modeling framework that is capable of real-time accurate electromagnetic responses of plasmonic nanogap structures for a wide range of geometry and material parameters. The main ingredients of the proposed method are: (i) the hybridizable discontinuous Galerkin method to numerically solve the equations governing electromagnetic wave propagation in dielectric and metallic media, (ii) a reference domain formulation of the time-harmonic Maxwell's equations to account for arbitrary geometry variations; and (iii) proper orthogonal decomposition and empirical interpolation techniques to construct an efficient reduced model. To demonstrate effectiveness of the models developed, we analyze geometry sensitivities and explore optimal designs of a 3D periodic coaxial nanogap structure.

\end{abstract}
  
\end{frontmatter}

The field of plasmonics \cite{maier2007plasmonics,ozbay2006plasmonics} studies the excitation of electrons in the conduction band of metallic nanostructures. This excitations can lead to resonances which result in the trapping of light in deep-subwavelength regions, leading to enormous near-field enhancements of the incident electromagnetic (EM) field. These regions of high-intensity EM field are typically observed near the corners, sharp features or cavities within metallic nanostructures. Moreover, the tight confinement and enhancement of the EM waves provide unique means for manipulating light and its interaction with metals, well beyond the diffraction limit. As a result, the development of plasmonic field enhancement structures has found applications in energy harvesting \cite{brongersma2016plasmonic},  sensing \cite{vspavckova2016optical}, plasmonic waveguiding  \cite{maier2005plasmonics} and near-field scanning microscopy \cite{novotny2011antennas}.

The ability to accurately model and simulate EM wave propagation for plasmonics requires capabilities that challenge traditional simulation techniques. Plasmonic phenomena involve the propagation of \mm and mm wavelengths through nanoscale apertures. It is thus essential to devise  simulation methods that are capable of resolving multiple length scales. From a manufacturing perspective, the fabrication of nanometer-wide gap structures is challenging, and minor geometric deviations from the nominal design may lead to significant alterations in the performance of the device. Furthermore, the optical behavior of the metals at the nanoscale is poorly characterized and the material constants in the models, which are determined experimentally, are subject to measurement errors. If our ultimate  goal is to develop simulation methods that can be effectively used in the design of plasmonic structures, it is of foremost importance that these methods incorporate fabrication tolerances as well as parameter uncertainties.

The goal of this paper is to combine an advanced numerical simulation method for the time-harmonic Maxwell's equations, the hybridizable discontinuous Galerkin (HDG) method, with a reduced order model (ROM) framework that will enable us to efficiently compute solutions for different combinations of the input parameters. These parameters may include material properties, frequency and direction of the incident light, as well as geometric dimensions. The ability to inexpensively obtain multiple solutions for different input parameters is critical to allow the effective exploration of the design space. The HDG method \cite{cockburn2008superconvergent,nguyen2009linearCD}, has been successfully applied to  acoustics and elastodynamics \cite{nguyen2011acoustic,saa2012binary}, as well as electromagnetics \cite{nguyen2011maxwell}. The HDG method is fully implicit in time, can be implemented on unstructured spatial discretizations, and can be made high-order accurate in both space and time, thus exhibiting low dissipation and dispersion. Compared to other DG methods \cite{huerta2013efficiency}, the HDG method results in a reduced number of globally coupled unknowns, thereby resulting in smaller linear systems, and exhibits optimal convergence rates for both the solution and the flux. Despite the computational advantage offered by the HDG method, the size of the resulting systems of equations, typically several hundred  thousand, is still too large to allow for the large number of solutions required in the design process. For this reason, we propose the construction of reduced order models (ROM) built upon the HDG method. These ROM allow for rapid and inexpensive queries of the outputs of interest for a range of geometry and problem parameters, effectively providing a parametrized solution of the problem.

The ROM of interest here are created by constructing a low-dimensional basis of the solution state space and seeking approximate solutions within this reduced space via Galerkin projection or other means. For this reason, they are often referred to as reduced basis (RB) methods \cite{hesthaven2016certified,quarteroni2011certified,quarteroni2015reduced}. Proper orthogonal decomposition (POD) \cite{sirovich1987turbulence} is commonly used to construct such low-dimensional bases. The starting point for POD is a set of solutions, or snapshots, that are obtained by numerically solving the forward HDG model for different values of the input parameters. Singular value decomposition is then employed to determine a low-dimensional basis such that the dominant information is retained by the basis vectors. POD has been used to construct ROM in computational fluid dynamics \cite{berkooz1993proper}, mechanical systems \cite{kerschen2005method} and electromagnetics \cite{albunni2008model}. In addition, alternative approaches to POD for creating ROM have been proposed. For instance, Arnoldi methods can be used to develop ROM for turbomachinery flows \cite{willcox2002arnoldi} and in recent years, Patera \etal \cite{grepl2007efficient,prud2002reliable,rozza2007reduced} introduced greedy sampling strategies to adaptively construct the RB, as well as rigorous \apo error estimators to certify the quality of the approximation and an offline/online strategy to improve computational efficiency. 
Other work has also been conducted to incorporate \apr error analysis \cite{fink1983error,maday2002global} of the approximation in ROM. The reduced RB representation can also be used to compute rapid and accurate approximations of functional outputs of parametrized PDEs \cite{noor1980reduced}.

Recently, RB methods have been used for the time-harmonic Maxwell's equations in combination with DG \cite{de2009reliable,kirchner2016maxwell,chen2010certified,pomplun2010reduced} and integral \cite{hesthaven2012certified,fares2011reduced} methods to successfully generate fast responses for complex electromagnetic devices, where the frequency and  material properties are treated as parameters. Furthermore, there has been considerable work to incorporate geometry as a parameter for electromagnetic applications, using either affine geometry transformations \cite{chen2011seamless,ganesh2012reduced,hess2013fast,w2014reduced} or parametrized meshes \cite{hammerschmidt2016reconstruction}.

The application of ROM techniques to linear PDEs with affine parameter dependence is fairly straightforward. However, the existence of nonlinearities and more complex parametric dependence, such as geometry parametrization, poses a severe challenge for RB. In that case,  the projection required to construct the reduced order model cannot be pre-computed and its cost scales as the dimension of the original solution space. In order to overcome this limitation, the empirical interpolation method (EIM) \cite{barrault2004empirical} and its discrete counterpart DEIM \cite{chaturantabut2010nonlinear} were proposed. These methods employ a greedy sampling strategy to generate an independent approximation to the nonlinear terms, consisting of a weighted combination of interpolation functions. Thus, replacing the nonlinear term with this interpolation approximation recovers an efficient computational strategy to perform the projection onto the low-dimensional space. In electromagnetics, the EIM has been leveraged to construct a ROM for 3D scattering problems \cite{ganesh2012reduced}. Alternatively, geometry parametrization has also been considered by directly interpolating the projection matrices that arise from the spatial discretization \cite{dyczij2009finite}, with applications to 2D and 3D waveguides \cite{burgard2013novel,burgard2014reduced}.

In order to pursue ROM within the HDG method, we first develop a reference domain formulation \cite{persson2009discontinuous} that renders a transformed set of time-harmonic Maxwell's equations for an arbitrary differentiable deformation mapping. We then combine the HDG discretization of the transformed equations with empirical interpolation techniques to recover a weak HDG formulation in terms of affine parametric expressions \cite{rozza2007reduced}. Finally, we employ POD to construct the RB for the ROM and the standard offline-online computational strategy that enables the evaluation of approximate solutions of large 3D problems in real-time. The main contribution of this article is the combination of existing numerical approaches such as HDG, EIM and ROM to address the simulation and design of novel plasmonic devices. The framework presented here is shown to dramatically reduce the computational cost of plasmonic simulations, therefore enabling the exploration of multiple material and geometry configurations.

This article is organized as follows. In Section \ref{sec:problem}, we introduce the equations and notation used throughout the paper. In Section \ref{sec:hdg}, we introduce the HDG method to solve time-harmonic Maxwell's equations on a reference domain and discuss its implementation. In Section \ref{sec:rom}, we describe the construction of a reduced order model for time-harmonic Maxwell's equations based on proper orthogonal decomposition and empirical interpolation. In Section \ref{sec:res}, we present numerical results to assess the performance of the proposed method. We finalize in Section \ref{sec:conc} by providing some concluding remarks.

\section{Governing equations}\label{sec:problem}

\subsection{Time-harmonic Maxwell's equations}
Maxwell's equations describe the propagation of the electric ${\mcal{E}}({\bx},{t})$ and magnetic ${\mcal{H}}({\bx},{t})$ fields. For monochromatic waves, we can express the electromagnetic fields using only their complex amplitudes, since the time dependence is assumed harmonic, that is $\mcal{E}(\bx,t) = \Re\lbrace \bE(\bx) \exp(-i\omega t)\rbrace$, for a given angular frequency $\omega$ and corresponding frequency $\fint  = \omega/(2\pi)$. For a linear, non-magnetic medium in the absence of impressed currents and charges, the frequency-domain Maxwell's equations for the complex electromagnetic amplitudes $\bE$ and $\bH$ are expressed as
\begin{equation*}
 \begin{aligned}
\nabla \times \bE -i\omega\bH &= 0 ,\\
\nabla \times \bH + i\omega\varepsilon \bE&= \sigma \bE,
\end{aligned}
\end{equation*}
where $\varepsilon$ is the relative permittivity of the medium and $\sigma$ is its conductivity.

\begin{figure}[h!]
 \centering
 \includegraphics[scale = .85]{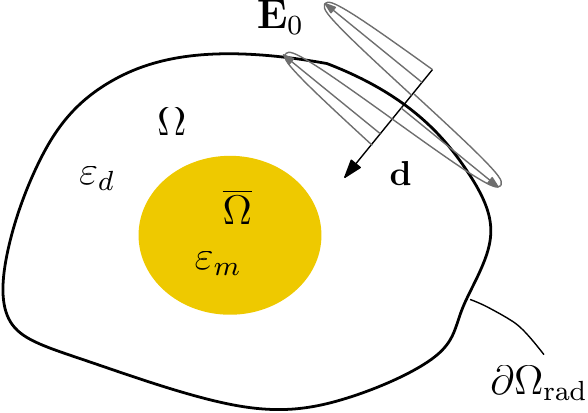}
 \caption{Metallic structure $\overline{\Omega}$ embedded in dielectric $\Omega$ illuminated by plane wave.}\label{fig:NLdomain}
\end{figure}

In this paper, we study the scenario where a plasmonic structure $\overline{\Omega}$ embedded in a dielectric medium $\Omega$ scatters an incident ${\bf p}$-polarized plane wave $\bE_0$ propagating in the ${\bf d}$-direction, that is ${\bf E}_0 = {\bf p}\exp(i\omega\sqrt{\varepsilon_d}\,{\bf d}\cdot \bx)$, see Fig \ref{fig:NLdomain}. The dielectric medium is characterized by a dielectric permittivity $\varepsilon_d$ and zero conductivity. The metal $\overline{\Omega}$ can be described with the complex Drude's permittivity \cite{drude1900elektronentheorie}, that is 
\begin{equation*}
\varepsilon_m = \varepsilon -\frac{\sigma}{i\omega} = \varepsilon-\dfrac{\omega_p^2}{\omega(\omega + i\gamma)}
\end{equation*} 
where $\varepsilon$ is permittivity due to the valence-band bound electrons, $\omega_p$ is the plasma frequency of the metal and $\gamma$ is the electron collision rate. This model assumes the electrons in the valence band are fully detached from the ions, and are only subject to electron-ion and electron-electron collisions. This model accounts for the dispersive and lossy nature of metals and is commonly known in literature as local model, since the motion of one electron is not coupled to that of its neighbors. For compactness, we use a spatially-varying permittivity $\varepsilon(\bx)$, defined as 
\begin{equation*}
\varepsilon(\bx) = \varepsilon_d \mathds{1}_\Omega + \varepsilon_m \mathds{1}_{\overline{\Omega}}.
\end{equation*}

Introducing the additional variable $\bV = i\omega\bH$, we arrive at the time-harmonic Maxwell's equations
\begin{equation}\label{eq:maxwell}
 \begin{aligned}
\nabla \times \bE -\bV &= 0 ,\\
\nabla \times \bV -\omega^2\varepsilon(\bx) \bE&= 0,
\end{aligned}
\end{equation}
defined in $\overline{\Omega} \cup \Omega$. The system above is complemented with boundary conditions
\begin{align*}
\label{eq:maxwellhydroV_BC}
&\bn\times\bE\times\bn  = \bE_\partial, \quad  \mbox{on }\partial{\Omega}_E,\\
&\bn\times\bV= \bn\times\bV_\partial,\quad  \mbox{on }\partial{\Omega}_V,\\
&\bn\times\bV +  i\omega\sqrt{\varepsilon_d}\,\bn\times\bE\times\bn = {\bf f}_0,\quad \mbox{on }\partial\Omega_{\rm{rad}}\,.
\end{align*}
with ${\bf f}_0 = \bn\times(\nabla\times {\bf E}_0) + i\omega\sqrt{\varepsilon_d}\bn\times\bE_0\times\bn $. The first (resp. second) equation corresponds to a boundary condition where the value of the tangential component of the electric (resp. magnetic) field is prescribed. Symmetries in the structure can be enforced by setting  $\bE_\partial = {\bf 0}$ ($\bV_\partial = {\bf 0}$) on selected boundaries, see Section \ref{sec:res}. The last equation imposes the radiation conditions required to truncate the infinite space. In particular, we employ Silver-M\"{u}ller conditions, which are first order absorbing boundary conditions \cite{sommerfeld1949partial,mur1981absorbing},

\subsection{Reference domain formulation}
\begin{figure}[h!]
 \centering
 \includegraphics[scale = .85]{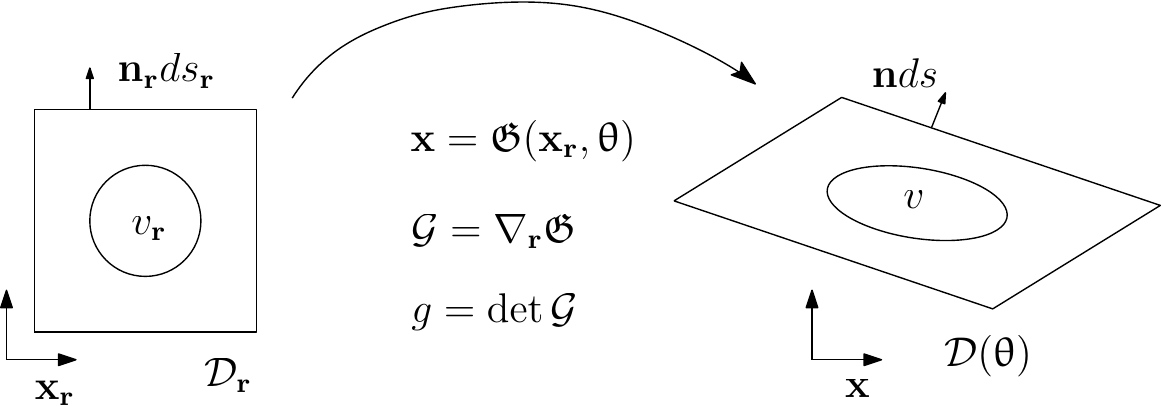}
 \caption{Deformation mapping between reference and physical domain.}\label{fig:mapping}
\end{figure}

In this section, we develop a reference domain formulation for the time-harmonic Maxwell's equations. Let us assume that the geometry of the physical domain, $\phdom(\upa)\in\mbb{R}^{n} $ with coordinates $\bx$, is parametrized by $\upa$ defined in a compact set. In addition, we consider a reference domain $\refdom\in\mbb{R}^{n}$ with coordinates $\refX$ which is mapped to the physical domain $\mcal{D}(\upa)$ by means of a one-to-one mapping given by a diffeomorphism $\bx = \mathfrak{G}(\refX,\upa)$, see Fig. \ref{fig:mapping}. Our goal is to pull Maxwell's equations in the physical domain \eqref{eq:maxwell}  back to the reference domain. The transformed equations in the reference domain are parametrized by $\upa$, but in return, they are defined and can be solved on a fixed, reference space, thus avoiding re-discretizing the computational domain for each new geometry, or value of the mapping parameters $\upa$. 

 The  deformation gradient of the mapping $\mathfrak{G}$ and its Jacobian are defined $\mcal{G} = \refV \mathfrak{G}$ and $g = \det \mcal{G}$, respectively. Note that $\nabla = \mcal{G}^{-T}\refV$. For simplicity, we shall assume deformations only occur within the vicinity of the metal, thus the outer boundaries of the computational domain -- where radiation conditions are prescribed -- remain fixed. For the plasmonic structures studied here, periodicity occurs in the $x-y$ directions and radiation conditions are imposed in the $z$ constant plane, thus the mapping can be easily decomposed to satisfy the required conditions, see \ref{app:map}. More general structures may either require more complex deformation mappings, or the use of perfectly matched layers (PMLs) \cite{berenger1994perfectly} to model radiation without the constraint of keeping the radiation boundaries fixed.
 
In order to obtain the transformed equations on the reference space $\refdom$, we consider, for illustration purposes, the second equation of \eqref{eq:maxwell} and integrate the curl term on a control volume $v\in\mcal{D}$. The application of Stokes' theorem and the relations $dv = gdv_\br$ and $\bn ds = g \mcal{G}^{-T} \refN ds_\br$ \cite{persson2009discontinuous} yields,
\begin{equation*}
\begin{split}
  \int_v \nabla \times { \bf V }dv & = -\int_s { \bf V} \times \bn ds = -\int_{s_\br} g{ \bf V} \times \lp\mcal{G}^{-T}\refN\rp ds_\br  =-\int_{s_\br} g\lp \mcal{G}^{-T} \lp\mcal{G}^{T}{ \bf V}\rp\rp \times \lp\mcal{G}^{-T}\refN \rp ds_\br \\&= -\int_{s_\br}\mcal{G} \lp \mcal{G}^{T}{ \bf V}\rp \times \refN ds_\br = \int_{v_\br} \mcal{G}\refV \times \lp \mcal{G}^{T}{ \bf V}\rp dv_\br\,,
\end{split}
\end{equation*}
The second term in transforms as
\begin{equation*}
\int_v \omega^2 \varepsilon(\bx) {\bf E}  dv = \int_{v_\br} \omega^2 \varepsilon(\bx_\br) {\bf E}  g dv_\br\,.
\end{equation*}
The boundary conditions are transformed similarly, for instance
\begin{equation*}
\begin{split}
   0 &=\int_s (\bE - \bE_\partial)\times\bn ds = \int_	{s_\br} g (\bE - \bE_\partial)\times(\mcal{G}^{-T} \refN)ds_\br = \int_{s_\br} \mcal{G} \lb\refN\times(\mcal{G}^T(\bE - \bE_\partial))\rb ds_\br \\
   &=  \int_{s_\br} \refN\times(\mcal{G}^T(\bE - \bE_\partial))ds_\br\,.
\end{split}
\end{equation*}
Finally, the transformed time-harmonic Maxwell's equations in $\overline{\Omega}_\br \cup \Omega_\br$ read
\begin{equation}\label{eq:maxwellALE}
\begin{aligned}
\refV \times \mG{{ \bf e}} - {{\bf v}} &=0,\\
\refV \times  \mG{{\bf v}}-\omega^2\varepsilon(\refX){\bf e} &= 0 ,  
\end{aligned}
\end{equation}
where a generic field ${\bf u}$ transforms as $ {\bf u}= g \mcal{G}^{-1} {\bf U}$ and $\mG = g^{-1} \mcal{G}^{T} \mcal{G}$. The system is complemented with boundary conditions 
 \begin{align*}
 &\refN \times(\mG\be)  = \refN\times(\mG\be_\partial), \quad  \mbox{on }\partial\Omega_{\br,E},\\
&\refN\times(\mG\bv)  = \refN\times(\mG\bv_\partial),\quad  \mbox{on }\partial\Omega_{\br,V},\\
&\refN\times(\mG\bv) + i\omega\sqrt{\varepsilon}\refN\times(\mG\be)\times\refN = {\bf f}_0, \quad \mbox{on }\partial\Omega_{\br,\rm{rad}}. 
 \end{align*}
Since deformations are only nonzero in the vicinity of the scatterer ($\mathfrak{G}{\restriction_{\partial\Omega_{\rm{rad}}}} = \mbox{Id}$), we have $\bV = \mG \bv$ and $\bE =\mG\be$ on $\partial \Omega_{\rm{rad}}$.

\section{HDG method for time-harmonic Maxwell's equations}\label{sec:hdg}

\subsection{Approximation spaces}

Following \cite{nguyen2011maxwell}, we denote by $\mcT{}$ a triangulation of disjoint regular elements $T$ that partition a domain $\mcal{D}\in\mbb{R}^3$. The set of element boundaries is then defined as $\partial \mcT{}:=\lbrace \partial T:\,T\in\mcT{} \rbrace$. For an arbitrary element $T\in\mcT{}$, $F = \partial T\cap \partial \mcal{D}$ is a boundary face if it has a nonzero 2D Lebesgue measure. Any pair of elements $T^+$ and $T^-$ share an interior face $F = \partial T^+ \cap \partial T^-$ if its 2D Lebesgue measure is nonzero. We finally denote  by $\mcal{E}_h^o$ and $\mcal{E}_h^\partial$ the set of interior and boundary faces respectively, and the total set of faces $\mcal{E}_h = \mcal{E}_h^o\cup \mcal{E}_h^\partial$.

Let $\bn^+$ and $\bn^-$ be the outward-pointing unit normal vectors on the neighboring elements $T^+,\,T^-$, respectively. We further use $\bu^\pm$ to denote the trace of $\bu$ on $F$ from the interior of $T^\pm$, where $\bu$ resides in $\bm L^2(\mcal{D})\equiv [L^2(\mcal{D})]^3$, with $L^2(\mcal{D})$ the space of square integrable functions on $\mcal{D}$. The jump $\llbracket\cdot\rrbracket$ for an interior face $F\in\mcal{E}_h^o$ is defined as
\begin{equation*}
 \llbracket \bu \odot \bn \rrbracket = \bu^+\odot\bn^+ + \bu^-\odot\bn^-,
\end{equation*}
and is single-valued for a boundary face $F\in\mcal{E}_h^\partial$ with outward normal $\bn$, that is
\begin{equation*}
 \llbracket \bu \odot \bn \rrbracket = \bu\odot\bn,
\end{equation*}
where the binary operation $\odot$ refers to either $\cdot$ or $\times$. The tangential $\bu^t$ and normal $\bu^n$ components of $\bu$, for which $\bu =\bu^t + \bu^n$, are then represented as
\begin{equation*}
 \bu^t:= \bn \times(\bu\times\bn),\qquad \bu^n := \bn(\bu\cdot\bn).
\end{equation*}

%Let $H^1(\mcal{D})$ denote the Hilbert space with $H^1(\mcal{D}) = \lbrace v\in L^2(\mcal{D}):\,\int_\mcal{D}\abs{\nabla v}^2 <\infty\rbrace$. We now introduce the curl-conforming space
%\begin{equation*}
% \bm H^{\curl}(\mcal{D}) = \lbrace \bu \in \bm L^2(\mcal{D}):\nabla\times \bu \in \bm L^2(\mcal{D}) \rbrace
%\end{equation*}
%with associated norm $\norm{\bu}^2_{ \bm H^{\curl}(\mcal{D})} = \int_{\mcal{D}} \abs{\bu}^2 + \abs{\nabla\times\bu}^2$, as well as the div-conforming space
%\begin{equation*}
% \bm H^{\dive}(\mcal{D}) = \lbrace \bu \in \bm L^2(\mcal{D}):\nabla\cdot \bu \in L^2(\mcal{D}) \rbrace
%\end{equation*}
%with associated norm $\norm{\bu}^2_{ \bm H^{\dive}(\mcal{D})} = \int_\mcal{D} \abs{\bu}^2 + \abs{\nabla\cdot\bu}^2$. 

Let $\mcal{P}^p(\mcal{D})$ denote the space of complex-valued polynomials of degree at most $p$ on $\mcal{D}$. We now introduce the following approximation spaces 
\begin{align*}
W_h &= \{w\in L^2(\mcal{D}) : w|_T \in \mcal{P}^{p}(T), \;\forall T\in \mathcal{T}_h\},\\
\bm W_h &= \{\bm \xi\in \bm L^2(\mcal{D}) : \bm \xi|_T \in \lb \mcal{P}^{p}(T) \rb^3,\; \forall T\in \mathcal{T}_h\},\\
M_h &= \{\mu\in  L^2(\mcal{E}_h)\,: \mu|_{F} \in \mcal{P}^{p}(F),\;\forall F\in \mcal{E}_h \},\\
\bm M_h &= \{\bm\mu\in \bm L^2(\mcal{E}_h)\,: \bm\mu|_{F} \in \mcal{P}^{p}(F)\bt_1 \oplus \mcal{P}^{p}(F)\bt_2,\;\forall F\in \mcal{E}_h \},
\end{align*}
where $\bt_1,\, \bt_2$ are vectors tangent to the face, thus including the curl-conforming nature of the solutions, since by construction $\bm\mu \in \bm M_h$ satisfies $\bm\mu = \bn\times( \bm\mu\times\bn ) = \mu_1 \bt_1 + \mu_2 \bt_2$. The tangent vectors can be defined in terms of $\bn = (n_1,n_2,n_3$) as ${\bf t}_1 =(-n_2/n_1,1,0)$ and ${\bf t}_2 =(-n_3/n_1,0,1)$. This definition assumes that $|n_1| \ge \max(|n_2|,|n_3|)$ but analogous expresions can be obtained when $|n_2| \ge \max(|n_1|,|n_3|)$ or $|n_3| \ge \max(|n_1|,|n_2|)$ to avoid division by a small number. Boundary conditions are included by setting $\bm M_h ({\bf u}_\partial) = \lbrace \bm \mu \in \bm M_h:\, \bn\times\bm\mu = \Pi {\bf u}_\partial\; \mbox{on } \partial \mcal{D} \rbrace$ and $ M_h ({ u}_\partial) = \lbrace  \mu \in  M_h:\, \mu = \Pi { u}_\partial\; \mbox{on } \partial \mcal{D} \rbrace$, where $\Pi {\bf u}_\partial$ (respectively, $\Pi u_\partial$) is the projection of ${\bf u}_\partial$ onto $\bm M_h$ (respectively, $u_\partial$ onto $M_h$).

Finally, we define the volume Hermitian inner products  as,
\begin{equation*}
 (\eta,\zeta)_{\mcT{}} := \sum_{T\in\mcT{}}(\eta,\zeta)_T,\qquad  (\bm\eta,\bm\zeta)_{\mcT{}} := \sum_{i = 1}^3(\eta_i,\zeta_i)_{\mcT{}},
\end{equation*}
and the surface Hermitian inner products as,
\begin{equation*}
 \langle\eta,\zeta\rangle_{\partial\mcT{}} := \sum_{T\in\mcT{}} \langle\eta,\zeta \rangle_{\partial T},\qquad   \langle\bm\eta,\bm\zeta \rangle_{\partial\mcT{}} := \sum_{i = 1}^3 \langle \bm\eta_i,\bm\zeta_i \rangle_{\partial\mcT{}}.
\end{equation*}

\subsection{Weak formulation}
In order to solve \eqref{eq:maxwellALE} on the reference domain $\Omega_\br\cup\overline{\Omega}_\br$, we first seek an approximation $({\bf v}_h,{\bf e}_h)\in \bm W_h\times\bm W_h$ to the transformed magnetic and electric fields $({\bf v},{\bf e})$. Furthermore, we also introduce a new variable $\widehat{\bf e}_h$ that approximates the tangential component of the transformed electric field at the element interfaces, that is $\refN \times ({\bf e}_h \times \refN)$. We finally introduce the approximation to the tangential component of the transformed magnetic field $\widehat{\bf v}_h$, and conservation is enforced by imposing continuity on the tangential component of $\mG \widehat{\bf v}_h$ across inter-element boundaries, that is $\llbracket \refN \times \mG\widehat{\bf v}_h \rrbracket = 0$ on $\mcal{E}_h^o$. 

The HDG method for the discretization of system \eqref{eq:maxwellALE} seeks a solution $({\bf v}_h,{\bf e}_h,\widehat{\bf e}_h) \in \bm W_h\times\bm W_h \times \bm M_h({\bf 0})$, such that the following system holds for all $(\bm \kappa,\bm \xi,\bm \mu) \in\bm W_h\times\bm W_h \times \bm M_h({\bf 0})$
\begin{align*}
(\bv_h,\bm \kappa)_{\mathcal{T}_h} - (\mG{\bf e}_h,\nabla\times\bm \kappa)_{\mathcal{T}_h}-\langle\mG\widehat{\bf e}_h,\bm \kappa\times \refN \rangle_{\partial \mathcal{T}_h}&= 0,\\
(\mG{\bf v}_h,\nabla\times \bm \xi)_{\mathcal{T}_h} + \langle\mG\widehat{\bf v}_h,\bm \xi\times \refN\rangle_{\mathcal{T}_h} - \omega^2(\varepsilon {\bf e}_h,\bm \xi)_{\mathcal{T}_h}  &=0 ,\\
-\langle \refN \times \mG\widehat{\bf v}_h ,\bm \mu\rangle_{\partial \mathcal{T}_h\backslash \partial\Omega_{\br,E}} + \langle\mG \widehat{\be}_h,\bm\mu \rangle_{\partial\Omega_{\br,E}} -i\omega\langle\sqrt{\varepsilon} \mG\widehat{\be}_h,\bm\mu\rangle_{\partial\Omega_{\rm{rad}}}& = \langle {\bf f} ,\bm \mu\rangle_{\partial\Omega_{\br}},
\end{align*}
where 
\begin{equation}\label{eq:bouflux_ref}
 {\bf f} = \mG{\be}_\partial{\restriction_{\partial\Omega_{\br,E}}} - \bn\times \mG\bv_\partial{\restriction_{\partial\Omega_{\br,V}}}  - {\bf f}_0{\restriction_{\partial\Omega_{\br,\rm{rad}}}}.
\end{equation}
and
\begin{equation*}
\mG\widehat{\bf v}_h = \mG{\bf v}_h + \tau_t\mG \lp {\bf e}_h - \widehat{\bf e}_h\rp \times \refN\,,
\end{equation*}
The global stabilization parameter $\tau_t$ is required to ensure the accuracy and stability of the HDG discretization, and defined as $\tau_t =\sqrt{\varepsilon}\omega$ \cite{nguyen2011maxwell}. Introducing the flux definition in the above system, we arrive to the final HDG system for the frequency-domain Maxwell's equations on a reference domain:
\begin{equation}\label{eq:hdgmaxwell_ALE}
 \begin{aligned}
(\bv_h,\bm \kappa)_{\mathcal{T}_h} - (\mG{\bf e}_h,\nabla\times\bm \kappa)_{\mathcal{T}_h}-\langle\mG\widehat{\bf e}_h,\bm \kappa\times \refN \rangle_{\partial \mathcal{T}_h}&= 0, \\
\begin{split}
(\mG{\bf v}_h,\nabla\times \bm \xi)_{\mathcal{T}_h} + \langle\mG{\bf v}_h ,\bm \xi\times\refN\rangle_{\mathcal{T}_h}- \omega^2(\varepsilon{\bf e}_h,\bm \xi)_{\mathcal{T}_h}& \\+ \langle\tau_t \mG({\bf e}_h - \widehat{\bf e}_h)\times \refN ,\bm \xi\times\refN\rangle_{\mathcal{T}_h}  &=0,
\end{split} \\
-\langle \refN\times \mG{\bv}_h + \tau_t \mG\be_h,\bm\mu \rangle_{\partial\mcal{T}_h \backslash \partial \Omega_{\br,E}} +\langle\widetilde{\tau}_t \mG\widehat{\be}_h,\bm\mu \rangle_{\partial\mcal{T}_h}&= \langle {\bf f},\bm\mu \rangle_{\partial\Omega_{\br}},
\end{aligned}
\end{equation}
where the stabilization constant $\widetilde{\tau}_t$ is given by
\begin{equation*}
\widetilde{\tau}_t = 
\begin{cases}
 \tau_t, & \mbox{on } \partial\mcal{T}_h \backslash \partial \Omega_{\br,E} \cup \partial \Omega_{\br,\rm{rad}} \\
 \tau_t - i\omega\sqrt{\varepsilon}, & \mbox{on } \partial\Omega_{\br,\rm{rad}} \\
  1, & \mbox{on } \partial\Omega_{\br,E} .\\
\end{cases}
\end{equation*}
Solving the weak formulation above is equivalent to solving the HDG discretization of \eqref{eq:maxwell} on the deformed domain $\mathfrak{G}(\overline{\Omega}_\br \cup \Omega_\br,\upa)$. However, we prefer the formulation in \eqref{eq:hdgmaxwell_ALE} since it allows us to work on a fixed parameter-independent domain and casts a clear strategy for model order reduction, as we shall describe below.

\subsection{Implementation}
We rewrite the weak formulation of system \eqref{eq:hdgmaxwell_ALE} as: find $({\bf v}_h,{\bf e}_h,\widehat{\bf e}_h) \in \bm W_h\times\bm W_h \times \bm M_h({\bf 0})$, such that
\begin{align}
\mathscr{A}({\bf v}_h,\bm \kappa) - \mathscr{B}(\mG{\bf e}_h,\bm \kappa)-\mathscr{C}(\mG\widehat{\bf e}_h,\bm \kappa)&= 0,\notag\\
\mathscr{B}(\mG{\bf v}_h,\bm \xi) + \mathscr{K}(\mG{\bf v}_h,\bm \xi) +\mathscr{D}(\mG {\bf e}_h,\bm \xi) -\omega^2 \mathscr{A}_\varepsilon({\bf e}_h,\bm \xi) -\mathscr{E}(\mG\widehat{\bf e}_h,\bm \xi) & = 0,\label{eq:hdg_weaksystemALE}\\
-\mathscr{R}(\mG{\bf v}_h,\bm\mu) -\mathscr{L}(\mG{\bf e}_h,\bm\mu) + \mathscr{M}(\mG\widehat{\bf e}_h,\bm\mu) & = \mathscr{F}(\bm\mu),\notag
\end{align}
holds for all $(\bm \kappa,\bm \xi,\bm\mu)\in \bm W_h \times \bm W_h \times \bm M_h({\bf 0})$. Here, the linear and  bilinear  forms are given by
\begin{equation*}\label{eq:bilinearForms}
\begin{array}{ll}
\mathscr{A}(\bv,\bm \kappa)= (\bv,\bm \kappa)_{\mathcal{T}_h}  ,&
\mathscr{A}_\varepsilon(\be,\bm \xi)= (\varepsilon \be,\bm \xi)_{\mathcal{T}_h}  ,\\
\mathscr{B}(\mG\be,\bm \kappa)= (\mG\be,\nabla\times \bm \kappa)_{\mathcal{T}_h},&
\mathscr{C}(\mG\widehat{\be},\bm \kappa)= \langle\mG\widehat{\be},\bm \kappa\times\bn\rangle_{\partial \mathcal{T}_h},\\
\mathscr{K}(\mG\bv,\bm \xi)= \langle\mG\bv,\bm \xi\times\bn\rangle_{\partial \mathcal{T}_h},&
\mathscr{D}(\mG\be,\bm \xi)= \langle \tau_t\mG\be,\bn\times \bm \xi \times\bn \rangle_{\partial \mathcal{T}_h},\\
\mathscr{E}(\mG\widehat{\be},\bm \xi) =\langle  \tau_t\mG\widehat{\be},\refN\times\bm \xi\times\refN\rangle_{\partial \mathcal{T}_h},&
\mathscr{R}(\mG\bv,\bm\mu) = \langle \mG\bv,\bm\mu\times\refN \rangle_{\partial \mathcal{T}_h},\\
\mathscr{L}(\mG\be,\bm\mu) = \langle \tau_t\mG\be,\bm\mu\rangle_{\partial \mathcal{T}_h},&
\mathscr{M}(\mG\widehat{\be},\bm\mu) = \langle \widetilde{\tau}_t\mG\widehat{\be},\bm\mu\rangle_{\partial \mathcal{T}_h},\\
\mathscr{F}(\bm\mu) = \mathscr{F}_E(\bm\mu) + \mathscr{F}_V(\bm\mu) + \mathscr{F}_{\rm{rad}}(\bm\mu)&
\end{array}
\end{equation*}
where the boundary form $\mathscr{F}$ arises from \eqref{eq:bouflux_ref}.

The discretization of the bilinear forms in \eqref{eq:hdg_weaksystemALE} gives rise to the system of equations
\begin{equation}\label{eq:hdgsystem}
\left[\begin{array}{ccc}
\mbb{A} & -\mbb{B} & -\mbb{C} \\
\mbb{B}+\mbb{K} & \mbb{D}-\omega^2\mbb{A}_\varepsilon   & -\mbb{E} \\
-\mbb{R}&  -\mbb{L} & \mbb{M} \\
\end{array}\right] \left[\begin{array}{c} \underline{\bv}  \\ \underline{\be}\\ \underline{\widehat{\be}}\end{array} \right] = \left[\begin{array}{c} 0 \\ 0\\ {\bf F} \end{array} \right]
\end{equation}
for the unknowns $\underline{\bv},\,\underline{\be},\,\underline{\widehat{\be}}$ corresponding to the degrees of freedom of the discretization $\mcal{T}_h$. In practice, however, the system \eqref{eq:hdgsystem} is never assembled. Instead, we invoke the discontinuity of the approximation spaces  and eliminate at the element level  $\underline{\bv}$ and $\underline{\be}$, also known as local variables, to obtain a system involving only the global variables $\underline{\widehat{\be}}$, hence reducing the size of the resulting linear system. The local variables may be rewritten as a function of the global variables using a Schur complement decomposition, namely
\begin{equation}\label{eq:hdg_local}
\left[\begin{array}{c} \underline{\bv} \\ \underline{\be}\end{array} \right] = \left[\begin{array}{cc}
\mbb{A} & -\mbb{B}  \\
\mbb{B}+\mbb{K} & \mbb{D}-\omega^2\mbb{A}_\varepsilon\\
\end{array}\right]^{-1}  \left[\begin{array}{cc}
\mbb{C} \\
\mbb{E}  \\
\end{array} \right]\underline{\widehat{\be}} = \mbb{Z}\, \underline{\widehat{\be}}\, ,
\end{equation}
where $\mbb{Z}$ has a sparse structure. That is, for a given element, the only nonzero columns correspond to the degrees of freedom of the faces of that element. The local linear systems \eqref{eq:hdg_local} are solved during the assembly process, and $\mbb{Z}$ is stored. Finally, the system involving only the global unknowns is given by 
\begin{equation}\label{eq:hdg_global}
\lp \mbb{M} + \left[\begin{array}{cc}
\mbb{R}&  \mbb{L}\\
\end{array}\right]\mbb{Z} \rp \underline{\widehat{\be}} ={\bf F} \,.
\end{equation}
This procedure characterizes the solution to \eqref{eq:maxwellALE} in terms of $\widehat{\be}_h$, which consists of only two components defined on the element faces. As a consequence, the HDG method exhibits less globally coupled degrees of freedom than other DG methods. Once the global variables  $\widehat{\be}_h$ have been computed, the local variables ${\bv}_h,\,{\be}_h$ can be inexpensively recovered in parallel using the pre-computed matrices $\mbb{Z}$ for each element independently. Finally, the approximate electromagnetic fields may be recovered from the transformed fields using the identities
\begin{equation}\label{eq:transformReference}
 \bV_h  = g^{-1}\mcal{G}\bv_h,\qquad \bE_h = g^{-1}\mcal{G}\be_h.
\end{equation}

\section{Reduced order modeling}\label{sec:rom}

The formulation and implementation introduced above describe the procedure for computing solutions to \eqref{eq:maxwellALE} for any combination of the parameter values $(\omega,\varepsilon,\upa)$. In order to describe the construction of ROM, we first rewrite the HDG system \eqref{eq:hdgsystem}  in a more compact form as follows: find a solution $\be_h^\rd := (\bv_h,\be_h,\widehat{\be}_h)$ in the $\mcal{N}^\rd$-dimensional approximation space $\bm W^\rd_h :=\bm W_h \times\bm W_h\times \bm M_h({\bf 0})$ satisfying 
\begin{equation}\label{eq:hdgsystem_compact}
 \mathscr{A}^\rd\lp \be^\rd_h,\bm \xi^\rd\,;(\omega,\varepsilon,\upa)\rp \be_h^\rd = \mathscr{F}^\rd \lp \bm \xi^\rd\,;(\omega,\varepsilon,\upa)\rp\;,
\end{equation}
for any $\bm \xi^\rd : = (\bm \kappa,\bm \xi,\bm\mu)\in{\bm W}^\rd_h$. 

Let us now assume that we are given a collection of solutions ${\be}^\rd_h \in {\bm W}^\rd_h$, or snapshots, computed at multiple parameter values through expressions \eqref{eq:hdg_local} and \eqref{eq:hdg_global}. In order to extract a low-dimensional representation of the parametric dependence, we apply POD and obtain a set of $N_{\max}$ orthonormalized basis functions $\snap_n\in {\bm W}^\rd_h,\;1\le n\le N_{\max}$ such that $(\snap_n,\snap_{n'})_{{\bm W}^\rd} = \delta_{nn'}$ for $1\le n,n'\le N_{\max}$, where the inner product $(\cdot,\cdot)_{{\bm W}^\rd}$ is defined as
\begin{equation*}
 ({\bm \xi}_1^\rd,{\bm \xi}_2^\rd)_{{\bm W}^\rd} = (\bm \kappa_1,\bm \kappa_2)_{\mcal{T}_h} + (\bm \xi_1,\bm \xi_2)_{\mcal{T}_h} +\langle \bm\mu_1,\bm\mu_2 \rangle_{\partial\mcal{T}_h}.
\end{equation*}
The choice $N_{\max}$ is made by monitoring the ratio between the energy in the model to the total energy of the snapshot matrix above using singular value information, see \cite{volkwein2011model} for more details.

The orthonormalized basis functions allow us to define an associated hierarchical POD space ${\bm W}^\rd_N$ as
\begin{equation*}
 {\bm W}^\rd_N = \mbox{span} \lbrace \snap_n,\,1\le n\le N \rbrace,\quad N = 1,\ldots,N_{\max}.
\end{equation*}

We now consider the Galerkin variational statement  \eqref{eq:hdgsystem_compact} but restricted to ${\bm W}^\rd_N \subset {\bm W}^\rd_h$ , that is: for an unseen parametric tuple $(\omega,\varepsilon,\upa)$, find an approximate solution ${\be}^\rd_N  \in {\bm W}^\rd_N$ satisfying
\begin{equation}\label{eq:hdgsystem_compact_noEI}
 \mathscr{A}^\rd\lp \be^\rd_N,\bm \xi^\rd\,;(\omega,\varepsilon,\upa)\rp \be_N^\rd = \mathscr{F}^\rd \lp \bm \xi^\rd\,;(\omega,\varepsilon,\upa)\rp\;,
\end{equation} 
for any $\bm \xi^\rd \in {\bm W}^\rd_N$. The above system of equations is low-dimensional and, in principle, we would expect that its evaluation for a new set of parameters is inexpensive. The main challenge, however, is that the geometry parameter $\upa$ enters in the definitions of  $\mathscr{A}^\rd$ and $\mathscr{F}^\rd$ through $\bm G$ in a \textit{non-affine} manner. This prevents us from isolating the parametric dependence and precomputing the required integrals required for the evaluation of $\mathscr{A}^\rd$ and $\mathscr{F}^\rd$. This means that solving system \eqref{eq:hdgsystem_compact_noEI} does not offer substantial computational advantages compared to the regular HDG system \eqref{eq:hdgsystem_compact}, since it still requires formation and assembly of the $\bm G$-dependent bilinear forms in \eqref{eq:hdg_weaksystemALE} for any new $\upa$. In order to circumvent this limitation, we resort to empirical interpolation techniques.

\subsection{Empirical interpolation}

% The main challenge is that affine parametric dependence is lost with respect to $\upa$, because even if $\mathfrak{G}$ were affine in $\upa$, the tensor $\bm G$ most surely is not. Hence, we must resort to advanced interpolation techniques to regain an affine expression for the weak formulation. 

A widespread technique in function interpolation is EIM \cite{barrault2004empirical}. The idea behind EIM is to approximate a non-affine parametrized function $f({\bf x},\upa)$ by a weighted combination of $\mcal{Q}$ orthogonal spatial functions $\bm\Phi = \lb \phi_1({\bf x}),\ldots,\phi_{\mcal{Q}}({\bf x})\rb$, that is $f({\bf x},\upa) \approx \bm\Phi({\bf x}) \bm\alpha(\upa)$ for a certain parameter-dependent coefficient vector $\bm\alpha(\upa)$. 

In order to apply the EIM to the bilinear forms \eqref{eq:hdg_weaksystemALE}, we first need to identify the non-affine parametrized functions to be interpolated, as well as the discrete set of spatial points ${\bf x}$ where the interpolants will be evaluated. In the spirit of keeping the main body of the article as clear as possible, we defer the implementation details of EIM to \ref{app:implementation}, and assume from this point onwards that both $\mathscr{A}^\rd$ and $\mathscr{F}^\rd$ may be approximated as
\begin{equation}
\begin{aligned}\label{eq:EIapproximations}
 \mathscr{A}^\rd\lp \cdot,\cdot\,;(\omega,\varepsilon,\upa)\rp &\approx   \sum_{q=1}^{\mcal{Q}} \alpha^{\mathscr{A}}_q(\upa) \mathscr{A}_q^\rd\lp \cdot,\cdot\,;(\omega,\varepsilon)\rp \\
  \mathscr{F}^\rd\lp \cdot\,;(\omega,\varepsilon,\upa)\rp &\approx  \sum_{q=1}^{\mcal{Q}} \alpha^{\mathscr{F}}_q(\upa) \mathscr{F}_q^\rd\lp \cdot\,;(\omega,\varepsilon)\rp
\end{aligned} 
\end{equation}
where the $\mathscr{A}_q^\rd,\,\mathscr{F}_q^\rd$ forms are affine in $\omega$ and $\varepsilon$.

If we now insert the approximations \eqref{eq:EIapproximations} into \eqref{eq:hdgsystem_compact_noEI}, we arrive at a linear system that is \textit{affine} in both the geometric parameters $\upa$ and the optical parameters $\omega,\,\varepsilon$ by construction, that is
\begin{equation*}
 \sum_{q=1}^{\mcal{Q}} \alpha^{\mathscr{A}}_q(\upa) \mathscr{A}_q^\rd\lp \be^\rd_N,\bm \xi^\rd\,;(\omega,\varepsilon)\rp \be_N^\rd = \sum_{q=1}^{\mcal{Q}} \alpha^{\mathscr{F}}_q(\upa) \mathscr{F}_q^\rd\lp \bm\xi^\rd \,;(\omega,\varepsilon)\rp \;,
\end{equation*} 
with solution ${\be}^\rd_N  \in {\bm W}^\rd_N$ for any ${\bm \xi}^\rd  \in {\bm W}^\rd_N$. 

Furthermore, for each new tuple $(\omega,\varepsilon,\upa)$ the system can be assembled extremely efficiently, since the dimension of the matrices is $N\times N$ due to the Galerkin projection using the low-dimensional approximation space $W^\rd_N$. The solution of this reduced order system are the trial coefficients $\lbrace \uplambda^\rd_n \rbrace_{n=1}^N$, that is ${\be}^\rd_N = \sum_{n=1}^N \uplambda^\rd_n \snapv_n$, enabling us to recover the approximate transformed electromagnetic fields as ${\be}_N^\rd = (\bv_N,\be_N,\widehat{\be}_N)$. The actual approximate electromagnetic fields $\bV_N,\,\bE_N$ are obtained from  $\bv_N,\,\be_N$ using \eqref{eq:transformReference} element-wise. Note that, for efficiency purposes, the non-affine function $g^{-1}\mcal{G}$ needed for the transformation can also be approximated with the EIM. Details concerning computational cost and implementation strategy are also described in \ref{app:implementation}.

\section{Numerical results}\label{sec:res}
\subsection{3D periodic coaxial nanogap structure}
\begin{figure}[h!]
\centering
\subcaptionbox{\label{fig:annularG10}}
[8cm]{\includegraphics[scale = 0.18]{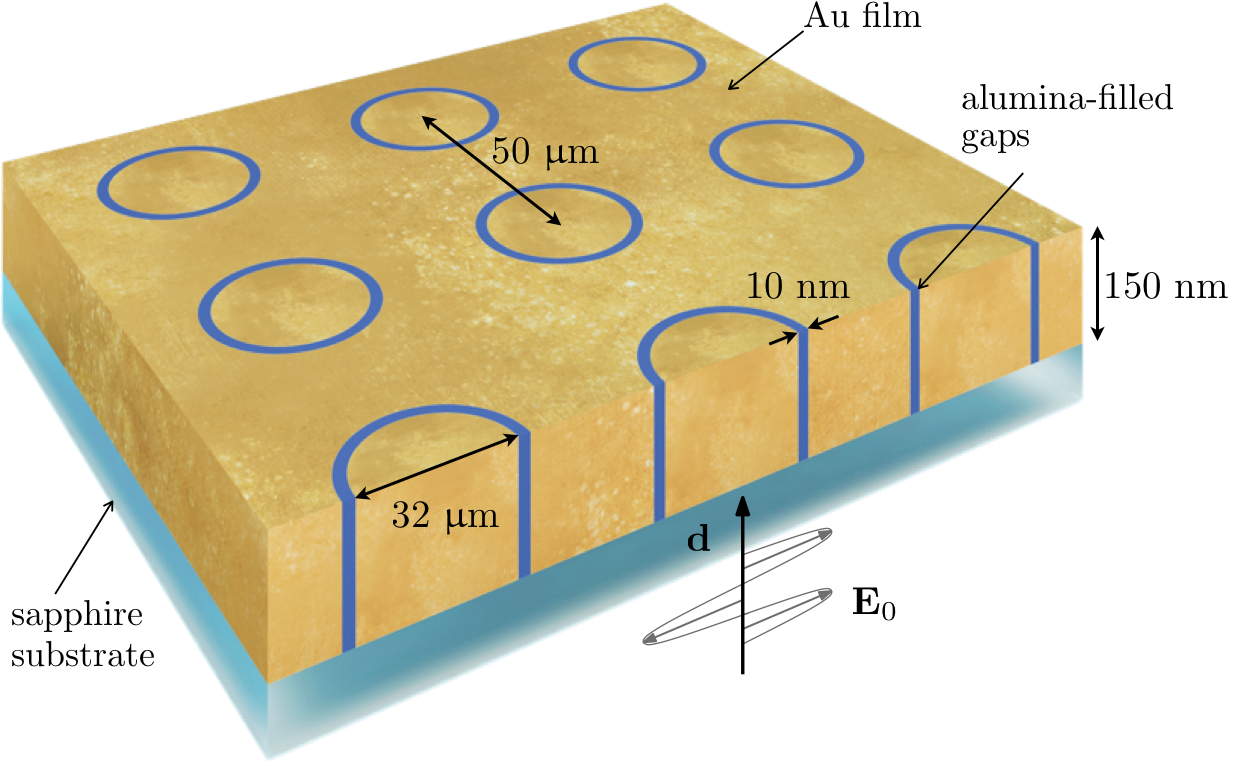}}
\hfill\subcaptionbox{\label{fig:annularFIR}}
[7cm]{\includegraphics[scale = 0.50]{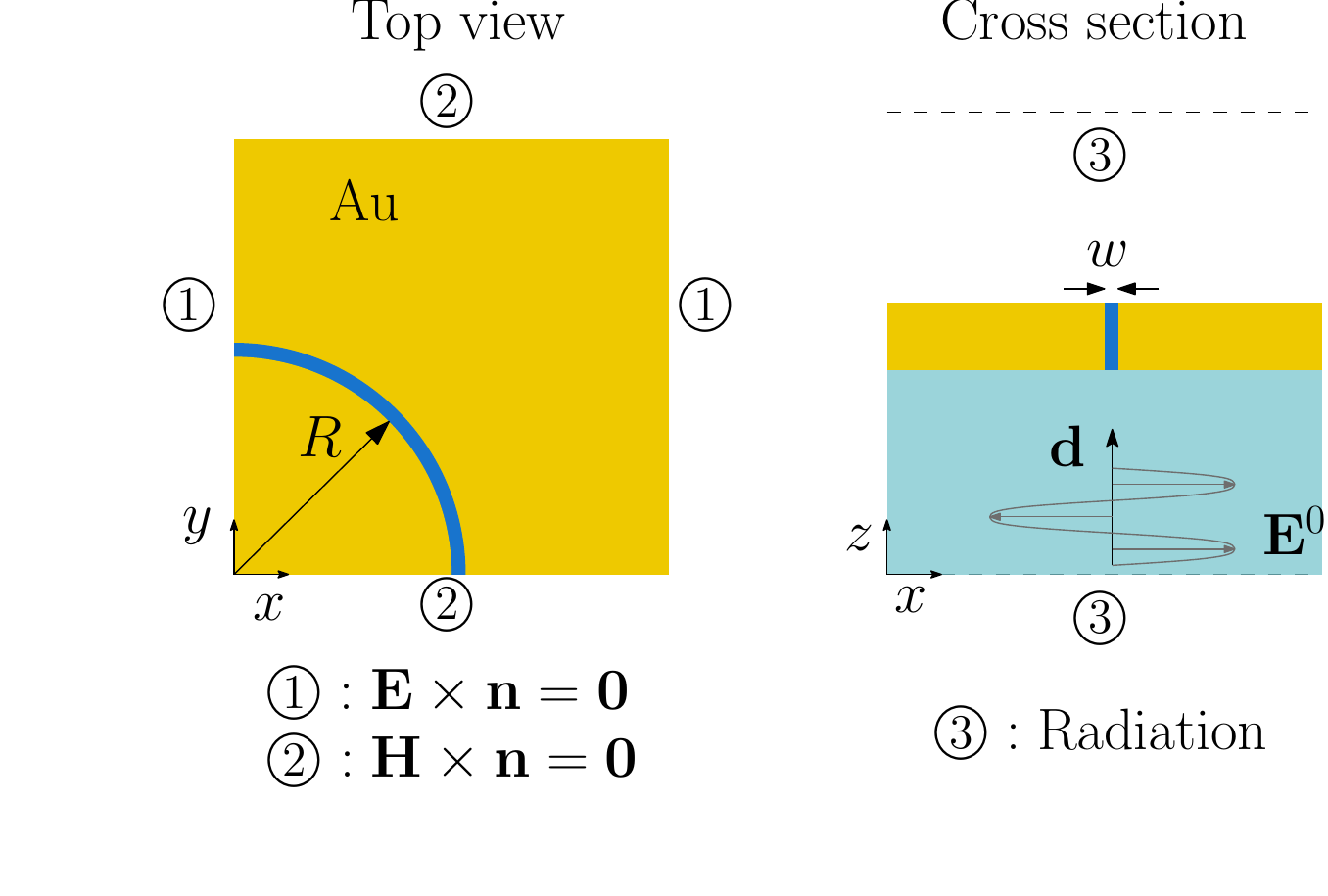}}
\caption{(a) Schematic diagram of thin gold film on silica substrate patterned with periodic square array of alumina gaps under plane wave THz illumination. (b)  Top and cross section of computational domain, with dimensions and boundary conditions.}\label{fig:annularAll}
\end{figure}

The first example consists of a 3D periodic annular structure shown in Fig. \ref{fig:annularG10}, which has been used for plasmonic sensing applications as well as fundamental studies of nanophotonics phenomena \cite{park2015nanogap,yoo2016high}. A 150 nm-thick gold film is deposited over a sapphire substrate, and the coaxial nanoapertures are filled with $\AO$. For this problem, we choose the diameter of the rings to be 32 microns with square periodicity of 50 \mm, and a gap width of 10 nm. 

% \begin{figure}[h!]
% \centering
% \includegraphics[scale = .2]{annularG10}
% \caption{Schematic diagram of thin gold film on silica substrate patterned with periodic square array of alumina gaps under plane wave THz illumination.}\label{fig:annularG10}
% \end{figure}

We illuminate the annular structure from below with an $x$-polarized plane wave, and for efficiency purposes we reduce the computational domain by exploiting the symmetries of the lattice, hence we only need to solve for one quadrant of the ring structure as indicated in Fig. \ref{fig:annularFIR}. Symmetries are imposed by setting $\bE\times\bn = \bm 0$ on the $x$-constant boundaries and $\bH\times\bn = \bm 0$ on the $y$-constant boundaries, whereas radiation conditions are prescribed on the $z$-constant boundaries. 

To analyze the performance of the structure, we focus on the electric field's $x$-component enhancement $\fie$ within the gap volume and the transmitted power $\pow$ through the structure, computed as
\begin{equation}
  \pow = \frac{\int_{A_1} \abs{\Re \lb \bE \times \bH^*\rb \cdot \bn} d{ A}}{\int_{A_0} \abs{\Re \lb \bE_0 \times \bH_0^*\rb \cdot \bn} d{ A}}\,,\qquad \qquad
\fie = \frac{\int_{\textnormal{gap}} \abs{\bE_x} dV}{\int_{\textnormal{gap}} \abs{\bE_{0,x}} dV}\,, 
\end{equation}
where $A_0$ is an arbitrary $xy$ plane below the gold film and $A_1$ an arbitrary $xy$ plane above the gold film.

The HDG discretization is composed of 1.3K hexahedral anisotropic elements, enabling us to accommodate the multiple length scales present in the structure with a reduced number of degrees of freedom, where a cubic representation of the geometry is employed to better represent the curved annular boundaries. Further details on the meshing strategies for these periodic coaxial nanostructures may be found in \cite{vidal2018hybridizable}. For the examples in this article, numerical accuracy is verified through grid convergence studies against a very fine mesh, until the relative error for the field enhancement is below 0.1\% at resonance.

\subsubsection{Material parameters}
One of the difficulties of simulating plasmonic devices is the determination of the permittivities for metals and dielectrics, since data that comes from measurements is often noisy and may exhibit significant variability. To that end, we develop a ROM considering the material properties as exploration parameters. The main objective is to show that, with few full model evaluations, we can construct a ROM that is capable of predicting inexpensively  the response of the plasmonic device for a range of material properties and frequencies. In this example, we  assume there are no geometry deformations, that is $\mathfrak{G} = \mbox{id}$.

Firstly, we define the parameters and their corresponding intervals of variability. We are interested in studying the field enhancement on the aperture volume of the coaxial structure introduced in Fig. \ref{fig:annularG10}, with frequencies ranging from 0.5 THz to 0.9 THz. The range of values used for the relevant parameters, along with the references wherefrom data was extracted, is summarized in Table \ref{tab:paramIntervals}. 
%For the optical constants of gold, we have considered the range of variation given by \cite{raether1980excitation,palik1998handbook}, even though it corresponds to a different frequency regime.

\begin{table}[h!]
\footnotesize
\begin{center}
  \renewcommand{\arraystretch}{1.2}% Spread rows out...
\begin{tabular}{cccc}

Parameter & Values  & Wavelength range ($\upmu$m) &References \\
\hline
$\hslash\,\widebar{\omega}_p$ [eV]  & $9.02\pm 0.18$ & < 12 &\cite{ordal1985optical,raether1980excitation,palik1998handbook}\\
$\hslash \,\widebar{\gamma}$ [eV] & $0.02678\pm0.007$ & < 12 & \cite{ordal1985optical,raether1980excitation,palik1998handbook}\\
$\epAO$ & 5 - 6 &  thickness-dependent &\cite{groner2002electrical}\\
$n_{\rm{sapphire}}$ & $3.07\pm 0.006$ & < 2 &\cite{grischkowsky1990far}\\

\end{tabular}
\end{center}
\caption{Variability ranges, interval of validity and references for parameters of structure in Fig. \ref{fig:annularG10}.}\label{tab:paramIntervals}
\end{table}

Secondly, the matrix of snapshots is obtained by sampling the 5-dimensional parameter space -- $\omega$ plus four material parameters in Table \ref{tab:paramIntervals}  -- and computing the full electromagnetic solution at 200 selected parameter values. The parameter values are chosen according to the Sobol sequence \cite{sobol1967distribution} in order to achieve a more uniform sampling in the high-dimensional space.

% \begin{figure}[h!]
% \centering
% \subcaptionbox{\label{fig:decayMaterial}}
% [7.5cm]{\includegraphics[scale = .35]{romdecay}}
% \hfill\subcaptionbox{ \label{fig:accuracyMaterial}}
% [7.5cm]{\includegraphics[scale = 0.35]{rommatonly}}
% \caption{(a) Decay of normalized singular values vs number of POD modes $N$. (b) Relative errors in field enhancement for test set parameters, evaluated for multiple POD sizes.}
% \end{figure}

\begin{figure}[h!]
\centering
\includegraphics[scale = .4]{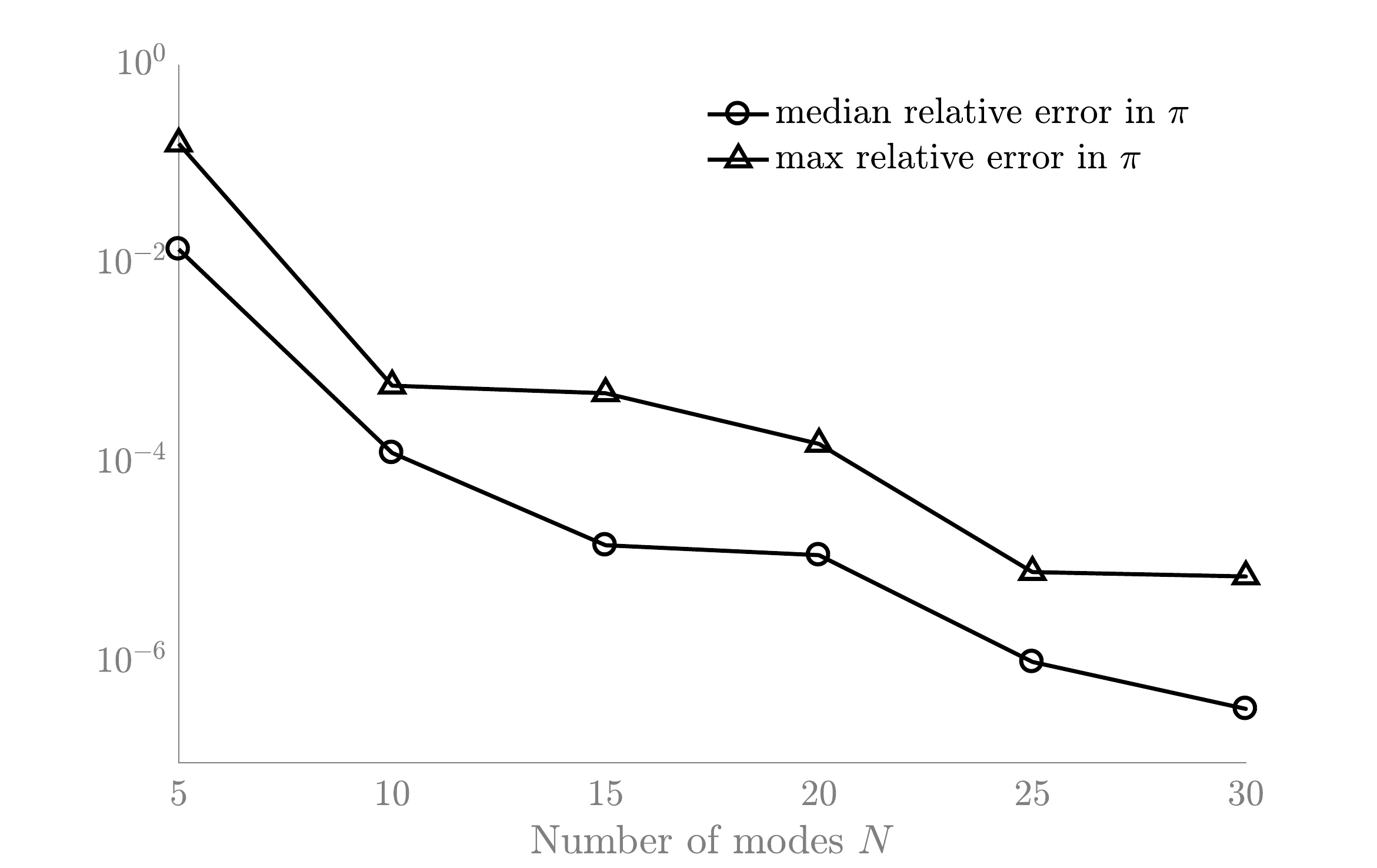}
\caption{Median and maximum relative error on field enhancement for the parameters in the test set, evaluated using 6 reduced models with increasing fidelity.}\label{fig:boxplot_materials}
\end{figure}

In order to assess the accuracy of the POD basis, we compute the HDG solution for 100 random parameter values -- referred to as the test set -- and evaluate the relative error committed  by the reduced order model on the test set of examples using up to $N=30$ modes in the basis. The relevant statistics of the distribution of relative errors clearly shows that the accuracy of the reduced model increases with the size of the basis $N$, see Fig. \ref{fig:boxplot_materials}. Furthermore, with only 10 modes 95\% of all testing examples are already below the threshold 0.1\%, which is generally deemed sufficient for most engineering applications. Note that the ROM does not converge exponentially for this problem, which may be due to the high dimensionality of the parameter space and the non-regularity of the quantity of interest.

The efficiency of the ROM is assessed by timing its online performance, more specifically the assembly, solution of the linear system and recovery of the local variables $(\bV_N,\bE_N)$ from the trial coefficients $\lbrace \uplambda^\rd_n \rbrace_{n=1}^N$. Time estimates are obtained averaging the wall time for 500 runs for each task using a single processor of a 512GB \Linux 12.04 machine with 32 AMD Opteron(tm) Processors 6320x15, and results are collected in Table \ref{tab:RBefficiency}. The advantages of the reduced order model are remarkable for this 3D problem, since we achieve an online cost reduction of $\sim 4$ orders of magnitude for the most expensive POD basis without compromising accuracy, as seen in Fig. \ref{fig:boxplot_materials}. As anticipated, the bulk of the computational cost to evaluate the POD basis is devoted to recovering the local variables required to evaluate the quantity of interest, since it involves operating with the high-dimensional POD basis functions. 
\begin{table}[h!]
\footnotesize
\begin{center}
  \renewcommand{\arraystretch}{1.2}% Spread rows out...
\begin{tabular}{lccc}
Model & Assembly  & Linear system &  $\bV_N,\,\bE_N$  \\
\hline
$N=5$ & 0.64 & 0.83 & 44.5\\
$N=10$ & 0.73 & 1.13 & 48.6\\
$N=15$ & 0.94 & 1.17 & 51.7\\
$N=20$ & 1.10 & 1.42 & 56.9\\
$N=25$ & 1.36 & 1.75 & 58.0\\
[1mm]
HDG & 7.6e5 & 1.8e6 & 8e3 \\

\end{tabular}
\end{center}
\caption{Times in s$\cdot10^{-4}$. Reduced model can be evaluated $\mcal{O}(10^4)$ times faster overall than HDG while maintaining accuracy.}\label{tab:RBefficiency}
\end{table}
This approach is particularly beneficial for large problems with parametric variability, where the model must produce accurate solutions in real-time for multiple queries.

The comparison above concerns only the online cost, but for completeness we shall now discuss the offline cost as well. The bulk of the offline cost is spent computing the HDG solutions that are later employed to construct the POD approximation space ${\bm W}^\rd_N$. Note however, that this solutions can be  easily computed in parallel. Since it is impossible to establish \apr the optimal number of snapshots required, we typically start with a small number of snapshots and then gradually enrich the solution space if the accuracy requirements on the test set are not met. All in all, creating a ROM for a 3D realistic problem is a computationally demanding task, hence it is only justified if the resulting ROM ought to be queried multiple times.

\subsubsection{Geometry deformations}
In this section, we demonstrate the effectivity of ROM for geometry deformations on the periodic coaxial structure introduced above. The objective is to show that, when the domain is subject to deformations, instead of solving Maxwell's equations on the deformed structure, it is desirable to solve the transformed Maxwell's equations \eqref{eq:maxwellALE} on the original reference structure in Fig \ref{fig:annularFIR}. 

% \begin{figure}[h!]
% \centering
% \includegraphics[scale = .35]{EIdecay}
% \caption{Normalized singular value decay of POD applied to nonaffine functions in Table \ref{tab:eim}, for model (R) in solid black and (RG) in dashed red.}\label{fig:decayPOD}
% \end{figure}

We analyze the effect of modifying the radius and gap width of the rings, with nominal values $R =16$ \mm and $w = 10$ nm. The modifications considered in this paper, for both radius and gap, are homogeneous with respect to the angle of the ring, hence $\upa=(R,w)$. The material parameters and gold optical constants are set to $\epAO = 5.5$, $n_{\rm{sapphire}} = 3.07$, $\hslash\omega_p = 9.02$ eV, $\hslash\gamma = 0.02678$ eV, $\varepsilon_\infty = 1$, and we study two different models: a $\pm30\%$ variation for the radius, that is $R \in [11.2,\,20.8]$ \mm, for a fixed gap $w = 10$ nm, denoted as $M_1$; and a $\pm20\%$ variation for both the radius and the gap, namely $R \in [12.8,\,19.2]$ \mm and $w \in [8,\,12]$ nm, denoted as $M_2$. Finally, we consider frequencies ranging from 0.3 THz to 0.9 THz. 

\begin{table}[h!]
\footnotesize
\begin{center}
  \renewcommand{\arraystretch}{1.2}% Spread rows out...
\begin{tabular}{ccccccc}

Model & $N$ &EIM coeff. & Assembly  & Linear system & $\bv_N,\,\be_N$ & $\bV_N,\,\bE_N$ \\
\hline
\multirow{3}{*}{$M_1$} & 25 & \multirow{3}{*}{21.78} & 1.83 & 0.21 & 2.97 & \multirow{3}{*}{4.90}\\
 &50 &  & 3.68 & 0.33 & 3.32 & \\
 &75 &  & 12.55 & 0.68 & 4.96 &\\
 &98 &  & 23.05 & 1.03 & 5.82 &\\
[2mm]
\multirow{3}{*}{$M_2$} & 25 & \multirow{3}{*}{21.85} & 2.07 & 0.18 & 2.60 & \multirow{3}{*}{4.85}\\
 & 50 & & 9.37 & 0.35 & 3.87 & \\
 & 71 &  & 20.25 & 0.66 & 4.89 &\\
  & 86 &  & 26.45 & 0.81 & 5.45 &\\
[2mm]
HDG & & -- &1.1e5 & 7.2e5 & 9.5e2 \\

\end{tabular}
\end{center}
\caption{Times in ms. Reduced model can be evaluated $\mcal{O}(10^3)$ times faster overall than HDG despite the EIM approximation.}\label{tab:RBefficiency_ALE}
\end{table}

The diffeomorphism $\mathfrak{G}$ prescribes variations of both the radius and the gap. The mapping requires continuity of the derivatives for the entire domain in order to ensure the face integrals in the weak formulation are well defined. To that end, we use $\mcal{C}^2$ cubic splines \cite{de1978practical} in the radial direction to construct the mapping, described in detail in \ref{app:map}.

% \begin{figure}[h!]
% \centering
% \includegraphics[scale = .4]{solutionsGeom}
% \caption{Relative error in field enhancement for test set parameters, evaluating both (R) and (RG) models at two POD sizes.}\label{fig:solutionsGeom}
% \end{figure}

%\begin{figure}[h!]
%\centering
%\subcaptionbox{Model $M_1$\label{fig:boxplot_R3}}
%[7.5cm]{\includegraphics[scale = .32]{boxplot_R3}}
%\hfill\subcaptionbox{Model $M_2$\label{fig:boxplot_GR2}}
%[7.5cm]{\includegraphics[scale = .32]{boxplot_GR2}}
%\caption{Quantiles and mean of the distribution of relative errors corresponding to the parameters in the test set, evaluated using 4 reduced models with increasing fidelity.}\label{fig:solutionsGeom}
%\end{figure}

% The POD basis for the radius (resp. radius-gap) model is formed by computing 350 (resp. 600) solutions of \eqref{eq:hdgmaxwell_ALE} at different $(\omega,\upa)$ values, and then compressed on a basis of 92 (resp. 86) modes. 
%The empirical interpolation can be performed very efficiently, since it only requires evaluating $\mathfrak{G}(\bx_\xi,\upa)$, its derivatives, and the subsequent nonaffine functions defined in Table \ref{tab:eim}. We then apply EIM separately for each function. The singular value decay associated with the POD of each nonaffine function is shown in Fig. \ref{fig:decayPOD} and we see that no more than 15 (resp. 30) modes are needed for any of the nonaffine functions.

\begin{figure}[h!]
\centering
\subcaptionbox{Field enhancement\label{fig:f_R}}
[7.5cm]{\includegraphics[scale = .35]{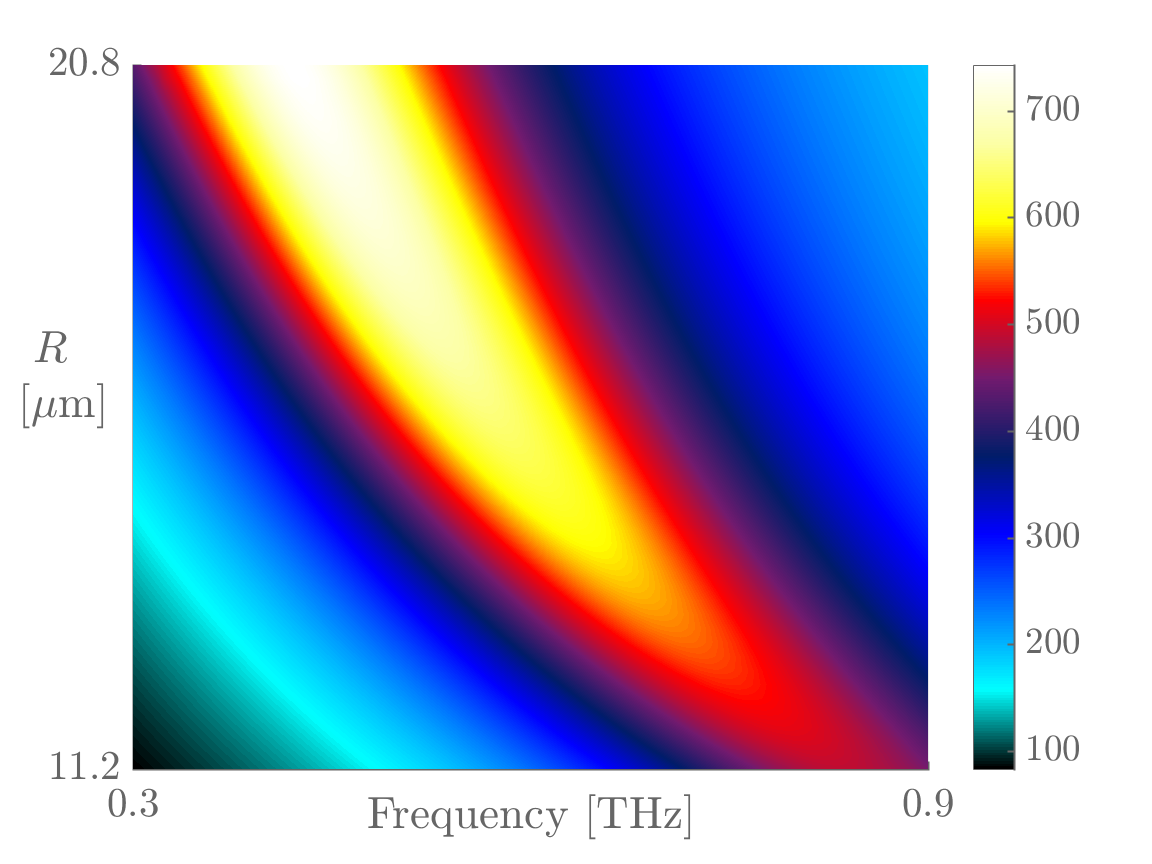}}
\hfill\subcaptionbox{Transmitted power (\%)\label{fig:tp_R}}
[7.5cm]{\includegraphics[scale = .35]{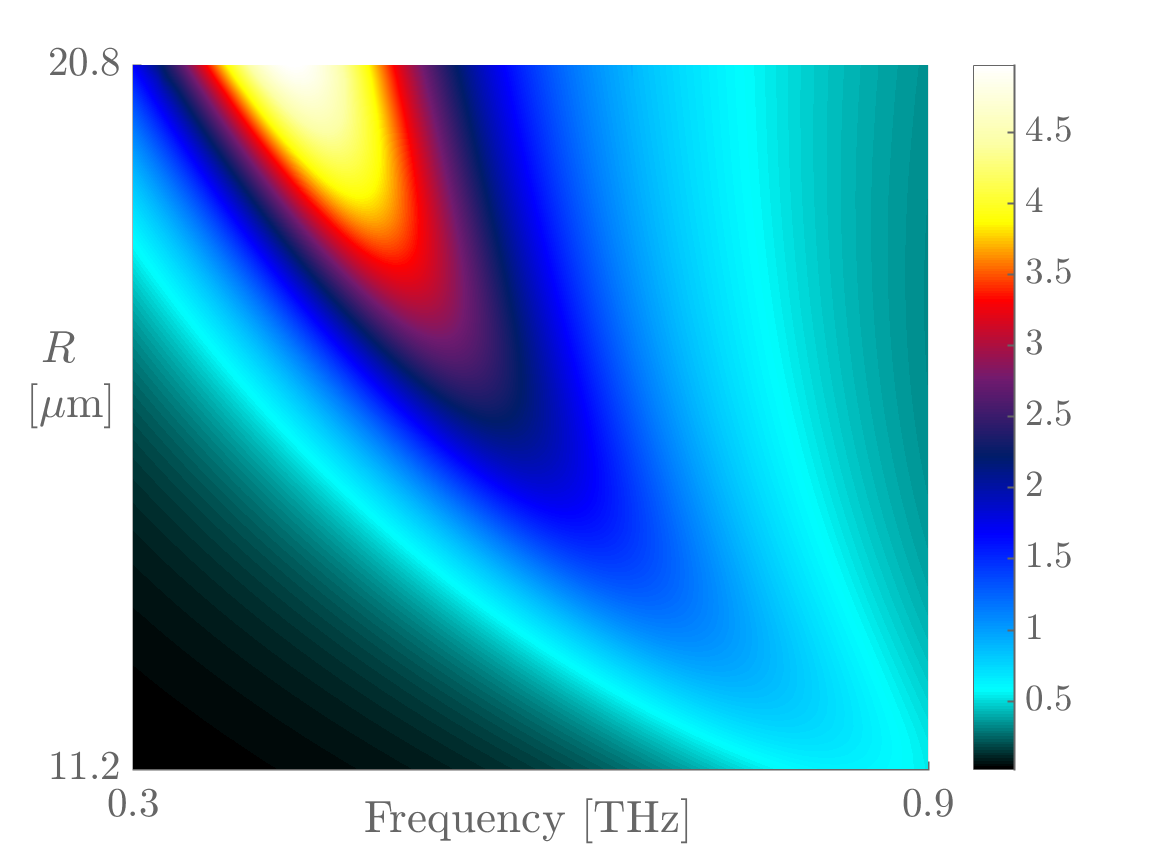}}
\caption{Resonant quantities maps for frequency and radius, evaluated with the $M_1$ reduced order model.}\label{fig:mapfR}
\end{figure}

The POD basis for $M_1$ (resp. $M_2$) is formed by computing 250 (resp. 400) solutions of \eqref{eq:hdgmaxwell_ALE} at different $(\omega,\upa)$ values, and then compressing them to form a basis of 98 (resp. 86) modes. Once the ROM is constructed, it can be efficiently queried for any valid combination of frequency and radius or radius-gap. We collect the wall time elapsed to evaluate the different pieces of the ROM. Recovering the empirical interpolation coefficients $\bm\alpha(\upa)$ is independent of the dimension of the ROM. Similarly as before, the bulk of the cost is incurred calculating the approximate transformed electromagnetic fields $\bv_N,\be_N$ from the trial coefficients, since it involves operating with the high-fidelity dimension $\mcal{N}^\rd$. Moreover, the reference domain formulation requires non negligible additional computations, since the approximate field variables $\bV_N,\bE_N$ need to be recovered through \eqref{eq:transformReference}. All in all, the benefits of employing a ROM to evaluate quantities of interest of the full EM simulation are remarkable, since a speedup of more than 3 orders of magnitude is achieved, even for the largest models, despite the overhead incurred by the empirical interpolation procedure, see Table \ref{tab:RBefficiency_ALE}. 
%As far as accuracy is concerned, we evaluate the field enhancement for a test set of parameters using the high-fidelity HDG model, and report the relevant statistics of the distribution of relative errors for both models in Fig. \ref{fig:solutionsGeom}. Although the increase in accuracy is modest as the ROM is enriched, relative errors below 0.1\% are achieved even with 50 modes.

Finally, we go one step further and show how ROM can be leveraged to achieve a deeper understanding of this structure. For instance, we can analyze the field enhancement and transmission profiles as a function of both the frequency and the radius of the annular aperture. For a fixed gap size, reducing the radius has a significant impact in the resonant quantities, more specifically blue-shifting the resonance, as shown in Fig. \ref{fig:mapfR}.

%In order to better visualize the effect of both the radius and the gap simultaneously for the (RG) model, we instead show the maximum field enhancement in Fig. \ref{fig:RG_FE} and the resonant frequency in Fig. \ref{fig:RG_f} as a function of these geometric parameters. As anticipated, reducing the gap width produces a greater enhancement of the field, due to superior localization, and redshifts the resonance. 

%\begin{figure}[h!]
%\centering
%\subcaptionbox{ \label{fig:f_R}}
%[5cm]{\includegraphics[scale = .33]{f_R}}
%\subcaptionbox{\label{fig:RG_FE}}
%[5cm]{\includegraphics[scale = 0.33]{RG_FE}}
%\subcaptionbox{\label{fig:RG_f}}
%[5cm]{\includegraphics[scale = 0.33]{RG_f}}
%\caption{Ring resonator parameter sweep results for: (a) Field enhancement profile for (R) model. (b) Maximum field enhancement for (RG) model. (c) Resonant frequency for (RG) model. }
%\end{figure}

\begin{figure}[h!]
\centering
\subcaptionbox{\label{fig:dgap_lam}}
[7.5cm]{\includegraphics[scale = .35]{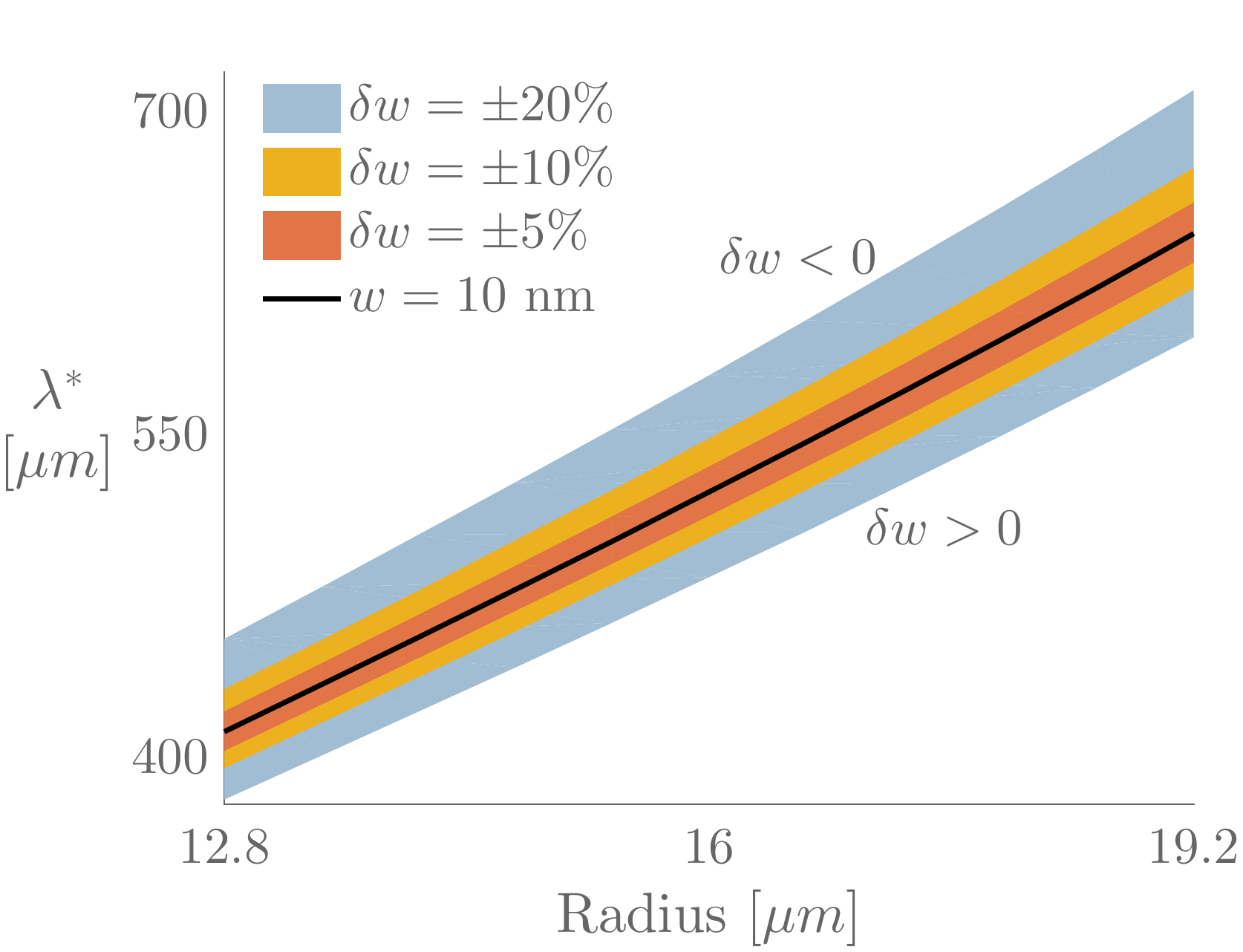}}
\hfill\subcaptionbox{\label{fig:dlam_dgap}}
[7.5cm]{\includegraphics[scale = .35]{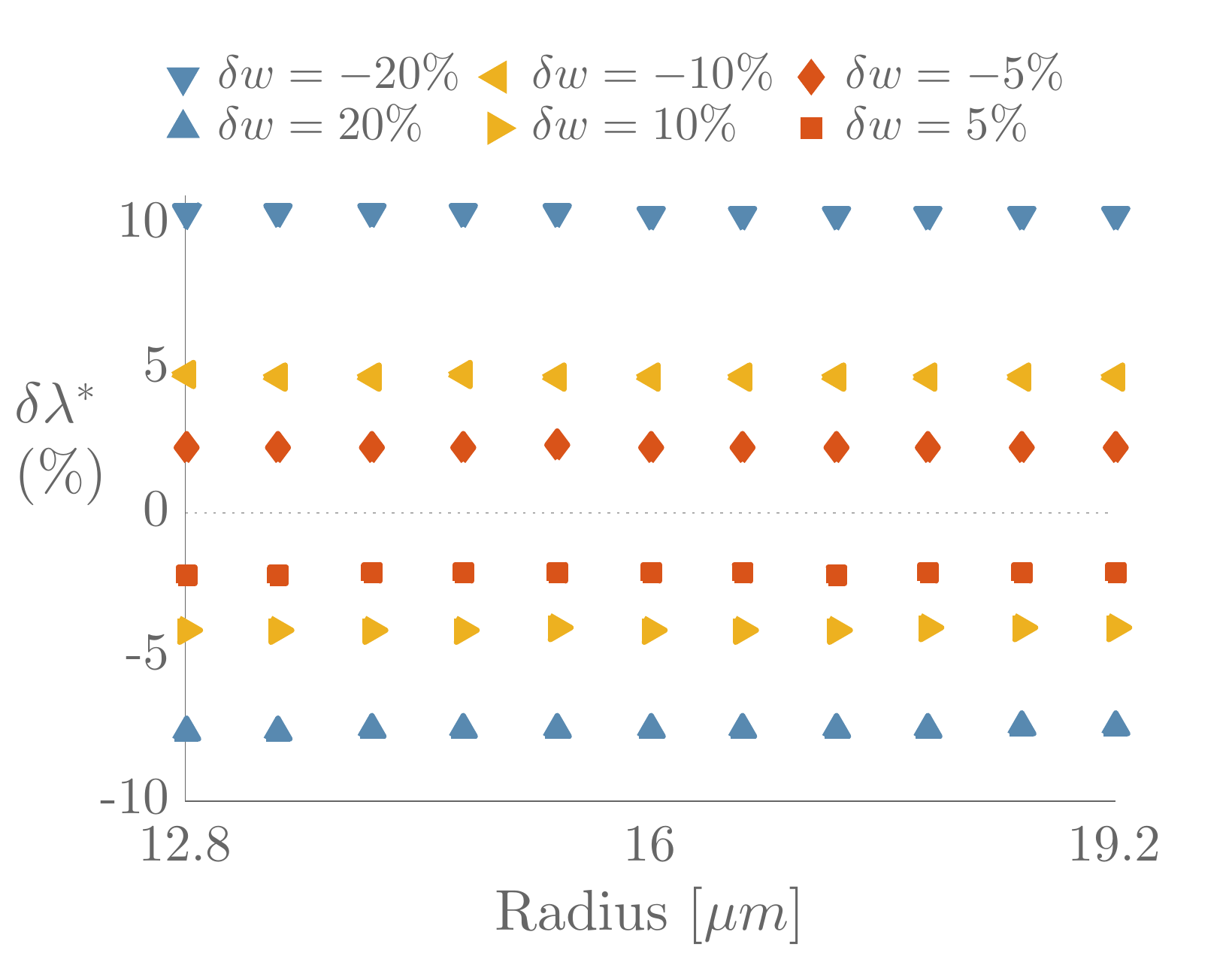}}
\hfill\subcaptionbox{\label{fig:drad_lam}}
[7.5cm]{\includegraphics[scale = .35]{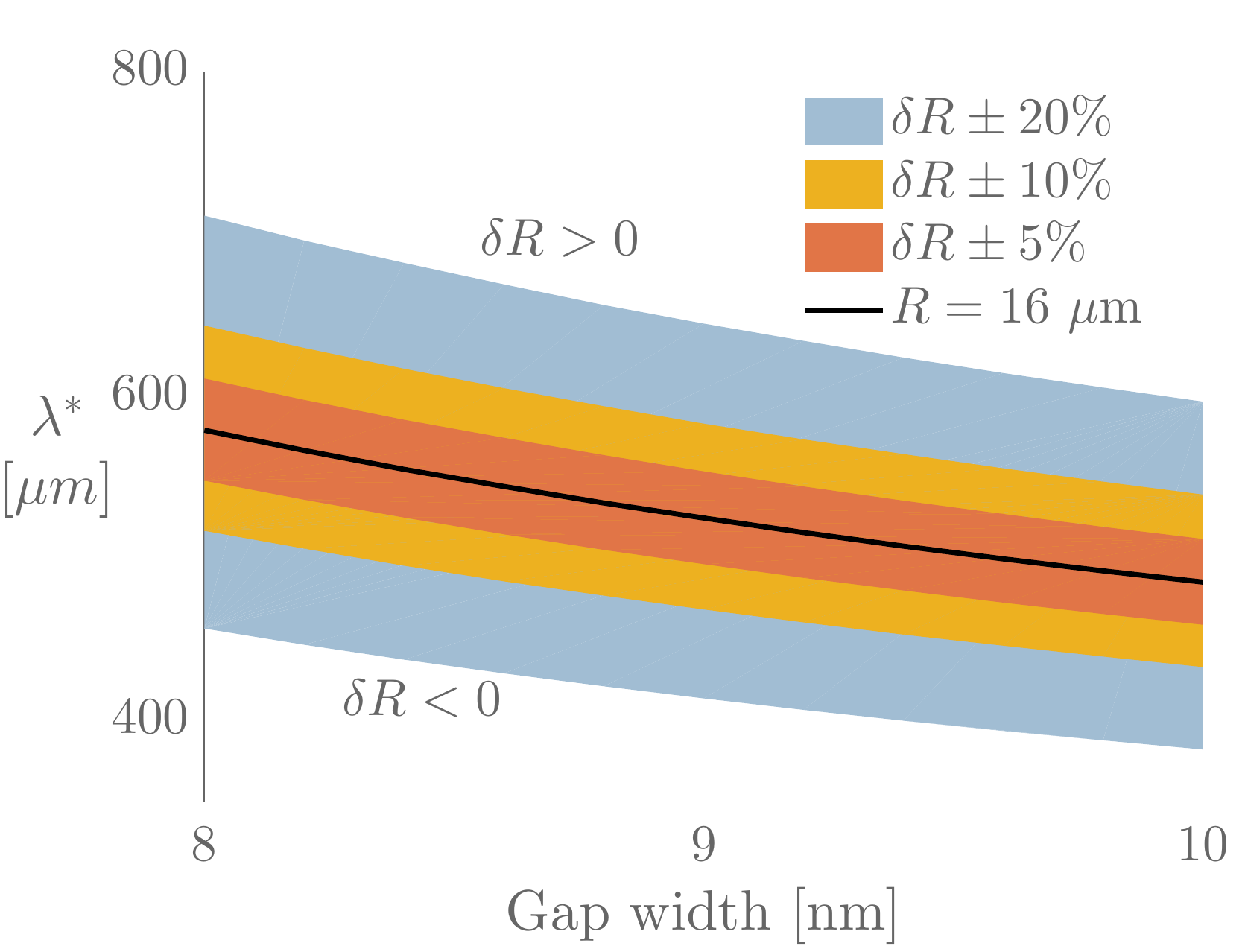}}
\hfill\subcaptionbox{\label{fig:dlam_drad}}
[7.5cm]{\includegraphics[scale = .35]{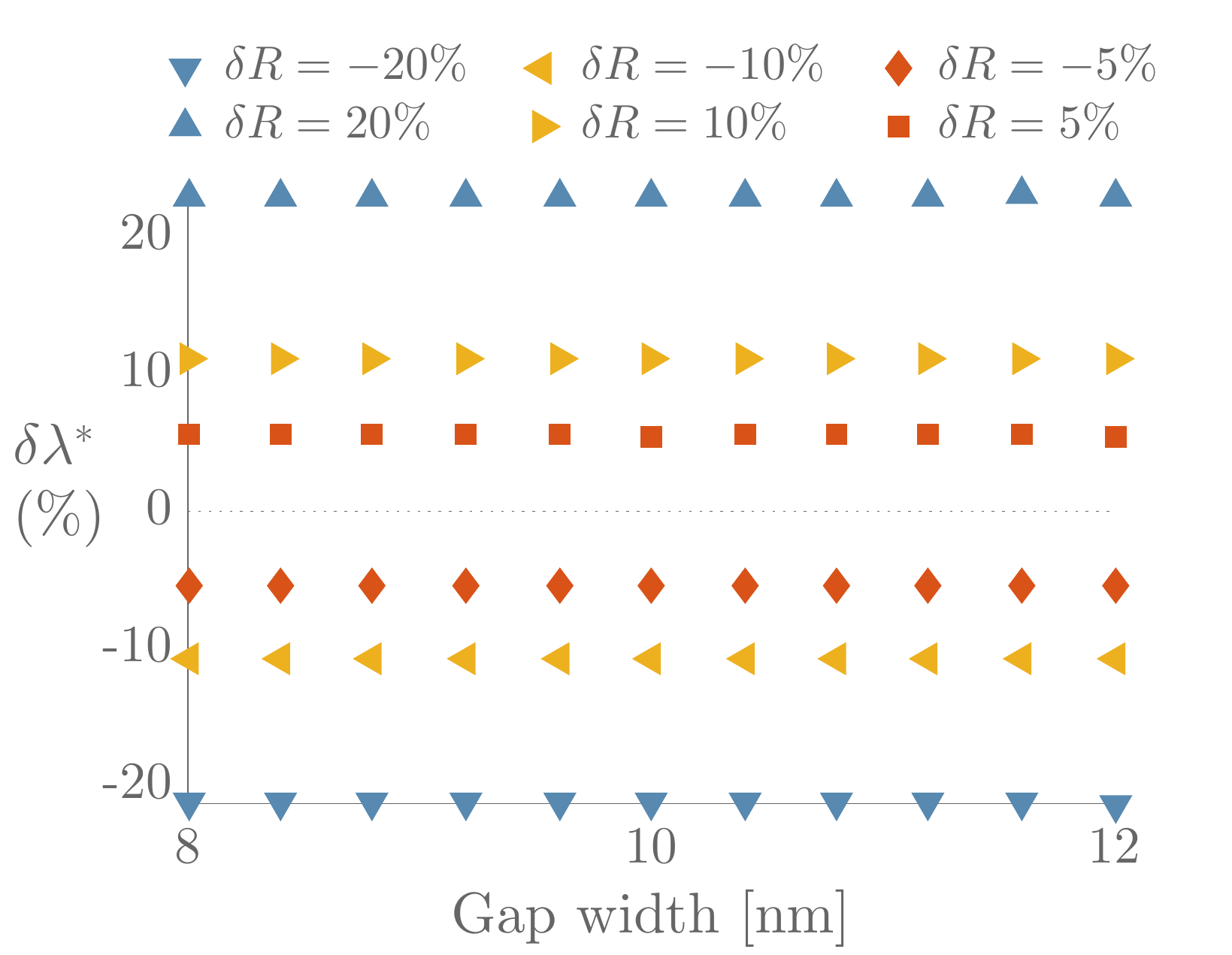}}
\caption{Absolute and relative sensitivities of resonant quantities on geometry modifications with $M_2$. (a)-(b) Gap variations and resonant wavelength. (c)-(d) Radius variations and resonant wavelength. }\label{fig:RG_all}
\end{figure}

Model $M_2$ gives rise to additional interpretations and results, since we can not only study the impact of both the radius and the gap width separately, but also their interactions. The value of the resonant wavelength $\lambda^*$ for the nominal gap width 10 nm as a function of the radius is depicted in Fig. \ref{fig:dgap_lam}. Additionally, the $\lambda^*$ dependence on the gap size for a fixed radius is shown in Fig. \ref{fig:drad_lam}. For both cases, we also provide intervals that correspond to the sensitivities of $\lambda^*$ with respect to $\pm5\%,\,\pm10\%$ and $\pm20\%$ relative variation of the either gap $\delta w$ or the radius $\delta R$. The ROM is essential to compute these results, since for each ring radius value and gap width a frequency sweep to detect the resonant quantities is required. 

Furthermore, additional relevant information may be extracted from the ROM. One such example is the relationship that maps gap relative variations $\delta w$ (resp. radius relative variations $\delta R$) to relative shifts in resonant wavelength $\delta\lambda^*$, see Fig. \ref{fig:dlam_dgap} (resp. Fig. \ref{fig:dlam_drad}). These mappings are obtained effortlessly once the ROM is available, and yet contribute to acquiring a more profound understanding of the device's behavior.

\subsection{Design of a concentric annular structure}
For this example, we will employ the ROM framework to investigate the annular structure with two concentric rings under several design criteria, and compare its performance with a single ring structure at low THz frequencies. Structures consisting of concentric coaxial nanogaps are expected to exhibit superior performance, since resonances are excited at distinct frequencies. More interestingly, the question of how to arrange the concentric apertures arises naturally, as different configurations may lead to significant changes in transmission. 

\begin{figure}[h!]
\centering
\includegraphics[scale = .25]{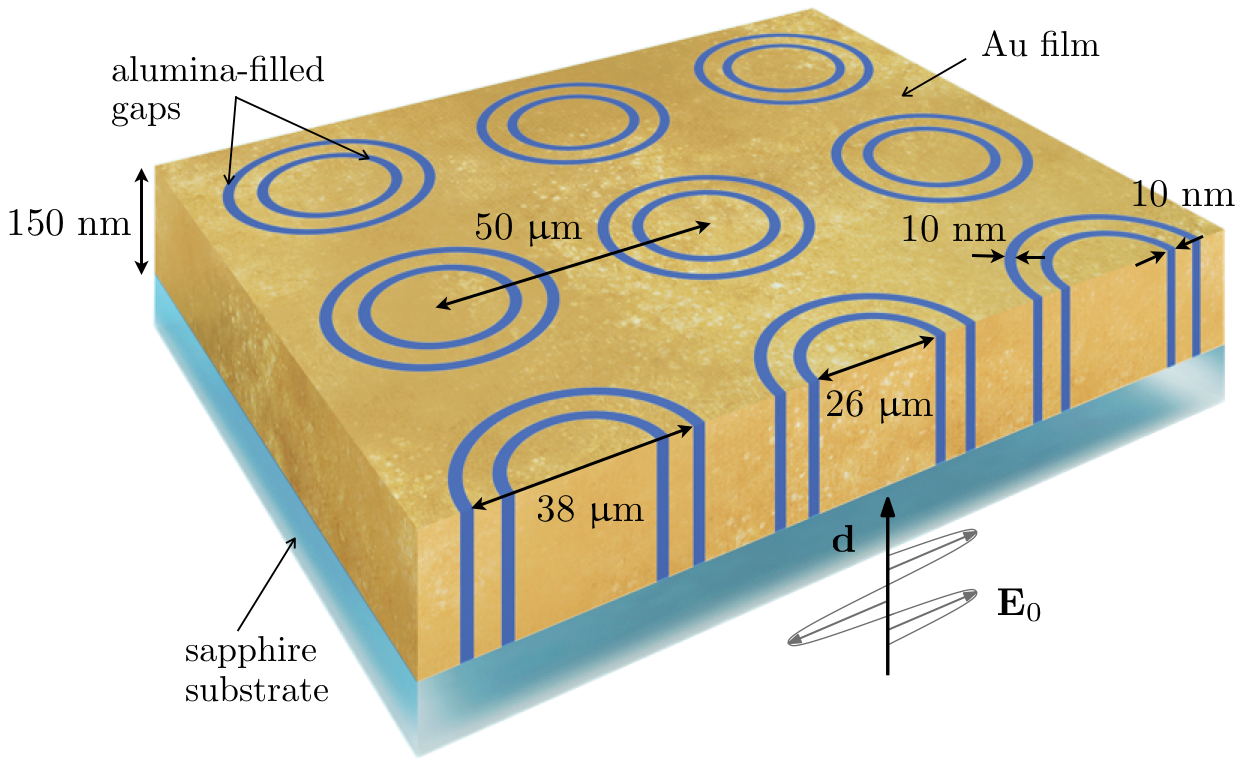}
\caption{Schematic diagram of thin gold film on sapphire substrate patterned with periodic square array of concentric alumina gaps under plane wave THz illumination.}\label{fig:annularFIR_CR}
\end{figure}

The reference structure is shown in Fig. \ref{fig:annularFIR_CR}, with ring diameters of 26 and 38 microns. The material parameters and gold optical constants are set to $\epAO = 5.5$, $n_{\rm{sapphire}} = 3.07$, $\hslash\omega_p = 9.02$ eV, $\hslash\gamma = 0.02678$ eV, $\varepsilon_\infty = 1$, and we consider frequencies in the range 0.3 THz to 0.9 THz. The gap widths are fixed to 10 nm, thus the only tunable geometric parameters are the radii of the rings $\upa = (R_1,R_2)$, set as $\pm 10\%$ of the nominal value, that is $R_1 \in [11.7,\,14.3]$ \mm and $R_2 \in [17.1,\,20.9]$ \mm. We then construct a surrogate of the high-fidelity model, using 650 snapshots computed over the three-dimensional parametric space formed by the $\omega$ and $\upa$, and results in a ROM with 74 modes.
%\begin{figure}[h!]
%\centering
%\includegraphics[scale = .35]{boxplot_2Gaps}
%\caption{Quantiles and mean of the relative errors distribution corresponding to the test set parameters, evaluated using 3 reduced models with increasing fidelity.}\label{fig:romCR}
%\end{figure}

% \begin{figure}[h!]
% \centering
% \subcaptionbox{\label{fig:annularFIR_CR}}
% [7.5cm]{\includegraphics[scale = .5]{annularFIR_CR}}
% \subcaptionbox{\label{fig:romCR}}
% [7.5cm]{\includegraphics[scale = .4]{romCR}}
% \caption{(a) Schematic of concentric annular resonator of diameters 26 \mm and 38 \mm in 150 nm gold film with sapphire substrate and array periodicity of 50 \mm. Computational domain is highlighted.  (b) }
% \end{figure}

In order to illustrate the potential of the concentric coaxial structure, we shall investigate three distinct objective functions to drive the design process, namely
\begin{equation*}
\mathcal{H}_1 = \max_{R_1,R_2} \varsigma,\qquad  \mathcal{H}_2 = \max_{R_1,R_2} E_{\fint}[\varsigma] - \sqrt{V_{\fint}[\varsigma]},\qquad \mathcal{H}_3 = \max_{R_1,R_2} E_{\fint}[\varsigma] - 10\sqrt{V_{\fint}[\varsigma]}
\end{equation*}
where $E_{\fint},\,V_{\fint}$ refer to the expectation and variance respectively, for frequencies ${\fint}$ in the range 0.4 THz to 0.7 THz. The objective function $\mathcal{H}_1$ seeks to maximize the transmitted power along the spectrum considered, whereas both $\mathcal{H}_2,\,\mathcal{H}_3$ target radii configurations that are robust to frequency variations (with distinct penalties) within the interval of interest. Fortunately, since the design space is only bidimensional it suffices to discretize it using a fine step size and compute the objective function invoking the ROM for all possible combinations of radii, thus avoiding the use of optimization algorithms. 

\begin{figure}[h!]
\centering
\subcaptionbox{\label{fig:objFunctionCoaxial}}
[15cm]{\includegraphics[scale = .34]{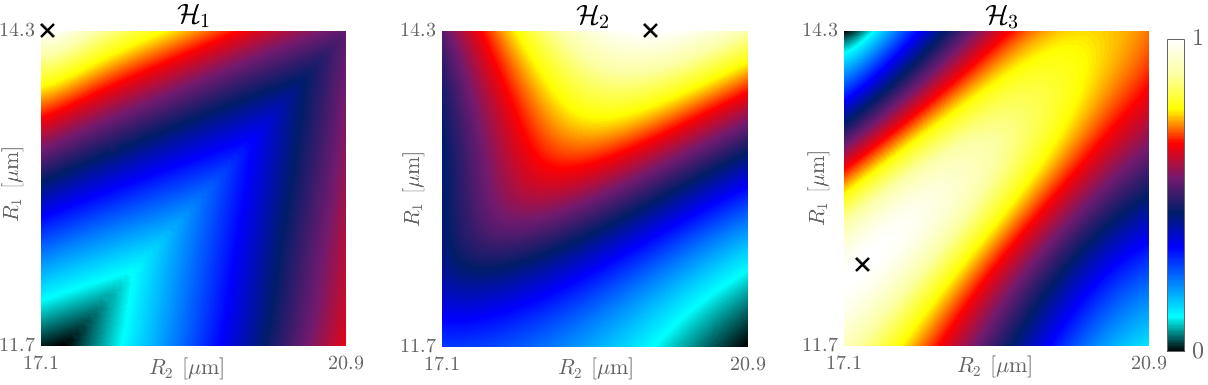}}
\hfill\subcaptionbox{\label{fig:resultsCoaxial}}
[15cm]{\includegraphics[scale = .36]{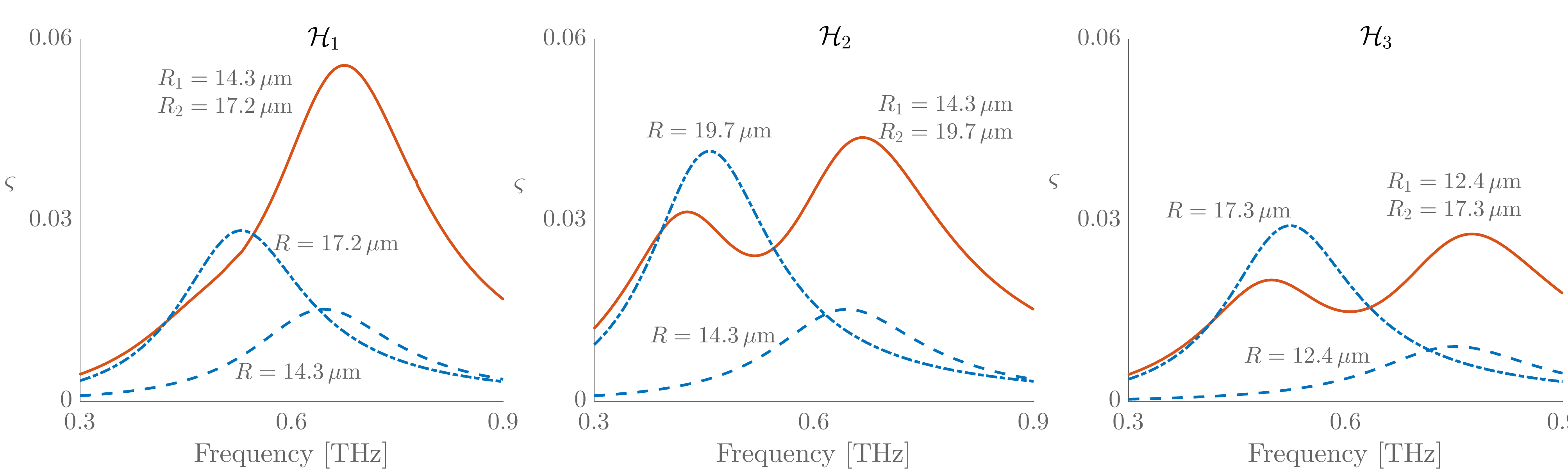}}
\caption{(a) Objective functions shown as surface plot for inner $R_1$ and outer $R_2$ radii values, normalized to $[0,\,1]$. Optimal configuration is shown with black cross. (b) Frequency-transmission profiles of optimal configurations for different objective functions.}
\end{figure}

The objective functions normalized between 0 and 1 along with the maximizing parameter configuration (marked with a cross) are represented in Fig. \ref{fig:objFunctionCoaxial}. For each objective function, the transmitted power $\varsigma$ spectrum for the optima radii (solid line), as well as the power transmitted by a structure with a single ring, for both the inner (dashed) and the outer (dash-dot) radius, are collected in Fig. \ref{fig:resultsCoaxial}.  Note that the response of the concentric ring structure differs significantly from the responses of the single ring configurations. Maximum transmission is attained when the concentric rings are separated by a small distance, whereby the transmission peaks that would correspond to each ring fuse into a single enhanced resonance. Moreover, the optimal configuration shown in Fig. \ref{fig:objFunctionCoaxial} (left) suggests that larger transmissions may be encountered if the separation is further reduced.

Conversely, if robust optima are sought, the ideal radii configurations result in increased separation, enabling the clear identification of the resonances corresponding to both annular apertures. The main changes between $\mathcal{H}_2$ and $\mathcal{H}_3$ is that the former prioritizes a larger average transmission on the interval of interest, whereas the latter renders a design less sensitive to frequency changes, at the expense of smaller average transmission. In both cases, the concentric ring structure provides increased transmission over a broader range of frequencies than that of the single ring.

\section{Conclusions} \label{sec:conc}
In this paper, we have developed a reduced order modeling framework for plasmonic coaxial nanogap structures based on the HDG method for Maxwell's equations. In order to account for the variability in the geometry of the plasmonic structures, we have proposed an alternative formulation of the HDG method for Maxwell's equations on a reference (parameter-independent) domain. The reference formulation, combined with proper orthogonal decomposition and empirical interpolation techniques, enables us to incorporate both material and geometric variability into the reduced order model. Furthermore, the ROM offline-online decomposition strategy allows us to compute solutions to Maxwell's equations inexpensively. The developed tool is advantageous for real design situations where multiple solutions for different values of the design parameters are required.

% Finally, we highlight several possible directions for future research. From the methodology perspective, an important shortcoming is the lack of \apo error bounds to quantify the error incurred by the surrogate model. The successive constraint method has been shown to produce inexpensive and sharp bounds for noncoercive problems \cite{huynh2007successive,huynh2010natural}. More recently, a novel approach based on a statistically modeling the mapping between inexpensive error bounds and the actual error using Gaussian processes has been proposed \cite{drohmann2015romes}. Incorporating error bounds may also be used towards driving the sampling process in a greedy fashion \cite{grepl2005posteriori,grepl2005reduced,veroy2003posteriori,veroy2005certified}. From the application perspective, it would be interesting to tackle the simulation and optimization of plasmonic devices with multiple geometry design variables, for example annular apertures with several concentric rings. The main caveat when accounting for more design parameters is always the construction of the ROM, since a greater number of snapshots is typically required to better explore the parameter space. Nonetheless, ROM can be a valid strategy to compute optimal designs of complex plasmonic structures.

\section*{Acknowledgements}
The authors acknowledge the support of AFOSR Grant No. FA9550-11-1-0141 and FA9550-12-0357. The authors also thank Prof. Anthony Patera and Prof. Sang-Hyun Oh for countless fruitful conversations, suggestions and comments.

\bibliography{mainbib}
\bibliographystyle{elsarticle-harv}

\appendix
\section{EIM on non-affine functions of HDG discretization} \label{app:implementation}
In this appendix, we provide extensive documentation on how to perform discrete empirical interpolation on the bilinear forms introduced in \eqref{eq:hdg_weaksystemALE}, thus enabling the approximation of the linear system in \eqref{eq:hdgsystem_compact_noEI} using the affine expression \eqref{eq:EIapproximations}. In addition, we review the computational strategy pursued to efficiently evaluate the reduced order model.

\subsection*{Discretization of bilinear forms}
Firstly, for the discretization of the forms in \eqref{eq:hdg_weaksystemALE}, we introduce $\varphi_i,\; 1\le i\le I$ to be the basis functions of $\mcal{P}^p(T)$ and $\phi_k,\, 1\le k \le K$ the basis functions of $\mcal{P}^p(F)$. Bearing in mind that the 3x3 deformation tensor $\mG$ is symmetric, we derive the contribution from an arbitrary element $T\in\mcal{T}_h$ to the global system. Furthermore, we shall make use of the identity $\bm\mu = \refN \times(\bm \mu \times\refN)$, for $\bm\mu \in \bm M_h({\bf 0})$  spanned by the reference-domain tangent vectors $\bt_1$ and $\bt_2$. 

The elemental matrices for element $T$ have the following  form:
\begin{alignat}{5}\label{eq:elementalHDG}
 \mbb{A}^t &= \begin{bmatrix}
              \mbb{A}_0^t & 0 &0\\ 0 &\mbb{A}_0^t &0\\ 0 &0 & \mbb{A}_0^t
             \end{bmatrix}\qquad  &\mbb{B}^t &= \begin{bmatrix}
              \mbb{B}_{11}^t & \mbb{B}_{12}^t &\mbb{B}_{13}^t\\ \mbb{B}_{21}^t & \mbb{B}_{22}^t & \mbb{B}_{23}^t\\ \mbb{B}_{31}^t & \mbb{B}_{32}^t & \mbb{B}_{33}^t
             \end{bmatrix} 
             \qquad  &\mbb{K}^t &= \begin{bmatrix}
              \mbb{K}_{11}^t & \mbb{K}_{12}^t &\mbb{K}_{13}^t\\ \mbb{K}_{21}^t & \mbb{K}_{22}^t & \mbb{K}_{23}^t\\ \mbb{K}_{31}^t & \mbb{K}_{32}^t & \mbb{K}_{33}^t
             \end{bmatrix} \notag\\
  \mbb{C}^t &= \begin{bmatrix}
              \mbb{C}_{11}^t & \mbb{C}_{12}^t \\ 
              \mbb{C}_{21}^t & \mbb{C}_{22}^t\\ 
              \mbb{C}_{31}^t & \mbb{C}_{32}^t
             \end{bmatrix} 
             \qquad &\mbb{D}^t &= \begin{bmatrix}
              \mbb{D}_{11}^t & \mbb{D}_{12}^t &\mbb{D}_{13}^t\\ \mbb{D}_{21}^t & \mbb{D}_{22}^t & \mbb{D}_{23}^t\\ \mbb{D}_{31}^t & \mbb{D}_{32}^t & \mbb{D}_{33}^t
             \end{bmatrix} \qquad 
             &\mbb{E}^t &= \begin{bmatrix}
              \mbb{E}_{11}^t & \mbb{E}_{12}^t \\ 
              \mbb{E}_{21}^t & \mbb{E}_{22}^t\\ 
              \mbb{E}_{31}^t & \mbb{E}_{32}^t
             \end{bmatrix}      
 \\
  \mbb{R}^t &= \begin{bmatrix}
              \mbb{R}_{11}^t & \mbb{R}_{12}^t &\mbb{R}_{13}^t\\ \mbb{R}_{21}^t & \mbb{R}_{22}^t & \mbb{R}_{23}^t
             \end{bmatrix} 
             \qquad  &\mbb{L}^t &= \begin{bmatrix}
              \mbb{L}_{11}^t & \mbb{L}_{12}^t &\mbb{L}_{13}^t\\ \mbb{L}_{21}^t & \mbb{L}_{22}^t & \mbb{L}_{23}^t
             \end{bmatrix}  \qquad 
             &\mbb{M}^t &= \begin{bmatrix}
              \mbb{M}_{11}^t & \mbb{M}_{12}^t \\ 
              \mbb{M}_{21}^t & \mbb{M}_{22}^t
             \end{bmatrix}                  \notag
\end{alignat}
Below, we provide expressions for the different subblocks for each elemental matrix, using for the volume $\varphi$ (resp. face $\phi$) basis functions $i/j$ (resp. $k/\ell$) as test/trial indices respectively, $1\le c,d \le 3$ for the dimension indices and $1\le a,b \le 2$ for the tangent vectors indices. The matrix $\mbb{A}^t$ consists of mass matrix subblocks $\mbb{A}_{0,ij}^t = (\varphi_i,\varphi_j)_T$ in the diagonal. The different components of the curl-convection matrices in $\mbb{B}^t$ are obtained as
\begin{align}\label{eq:EIM_B}
 \mbb{B}^t_{1d,ij} & = (\partial_z\varphi_i,\mG_{2d}\,\varphi_j)_T - (\partial_y\varphi_i,\mG_{3D}\,\varphi_j)_T\;, \notag\\
  \mbb{B}^t_{2d,ij} & = (\partial_x\varphi_i,\mG_{3D}\,\varphi_j)_T - (\partial_z\varphi_i,\mG_{1d}\,\varphi_j)_T\;, \\
  \mbb{B}^t_{3D,ij} & = (\partial_y\varphi_i,\mG_{1d}\,\varphi_j)_T - (\partial_x\varphi_i,\mG_{2d}\,\varphi_j)_T. \notag
\end{align}
The subblocks for $\mbb{K}^t$ and $\mbb{D}^t$ are $\mbb{K}^t_{cd,ij} = \langle \kappa_{cd} \varphi_i,\varphi_j\rangle_{\partial T}$ and $\mbb{D}^t_{cd,ij} = \langle \delta_{cd} \varphi_i,\varphi_j\rangle_{\partial T}$, where $\kappa_{cd},\,\delta_{cd}$ are given by
\begin{align}
 \kappa_{1d} & = \mG_{3D} n_2 - \mG_{2d}n_3\;, \quad \kappa_{2d}  = \mG_{1d} n_3 - \mG_{3D}n_1\;, \quad  \kappa_{3D}  = \mG_{2d} n_1 - \mG_{1d}n_2\;,\label{eq:EIM_K}\\  
  \delta_{1d} & = \mG_{1d} (1-n_1^2)- n_1n_3 \mG_{3D} - n_1n_2\mG_{2d}\;,\notag\\
  \delta_{2d} &= \mG_{2d} (1-n_2^2)- n_1n_2 \mG_{1d} - n_2n_3\mG_{3D}\;, \label{eq:EIM_D}\\
  \delta_{3D} &= \mG_{3D} (1-n_3^2)- n_1n_3 \mG_{1d} - n_2n_3\mG_{2d}\;.\notag
\end{align}
The submatrices for $\mbb{C}^t,\,\mbb{E}^t$ are given by $\mbb{C}^t_{cb,i\ell} = \langle \gamma_{cb} \varphi_i,\phi_\ell\rangle_{\partial T},\;\mbb{E}^t_{cb,i\ell} = \langle \epsilon_{cb} \varphi_i,\phi_\ell\rangle_{\partial T}$. For simplicity, we introduce the modified tangent vectors $\widehat{\bf t}_b = \mG {\bf t}_b$, and define $\gamma_{cb},\,\epsilon_{cb}$ as
\begin{align}
 \gamma_{1b} & = \widehat{t}_{b3} n_2 - \widehat{t}_{b2}n_3\;, \quad \gamma_{2b}  = \widehat{t}_{b1} n_3 - \widehat{t}_{b3}n_1\;, \quad  \gamma_{3b}  = \widehat{t}_{b2} n_1 - \widehat{t}_{b1}n_2\;,\label{eq:EIM_C}\\  
  \epsilon_{1b} & = \widehat{t}_{b1} (1-n_1^2)- n_1n_3 \widehat{t}_{b3} - n_1n_2\widehat{t}_{b2}\;,\notag\\
  \epsilon_{2b} &= \widehat{t}_{b2} (1-n_2^2)- n_1n_2 \widehat{t}_{b1} - n_2n_3\widehat{t}_{b3}\;,  \label{eq:EIM_E}\\
  \epsilon_{3b} &= \widehat{t}_{b3} (1-n_3^2)- n_1n_3 \widehat{t}_{b1} - n_2n_3\widehat{t}_{b2}\;. \notag
\end{align}
In addition, the $\mbb{R}^t$ components are given by $\mbb{R}^t_{ad,kj} = \langle [\mG({\bf t}_a\times\refN)]_d \phi_k,\varphi_j \rangle_{\partial T}$ and the $\mbb{L}^t$ submatrices by $\mbb{L}^t_{ad,kj} = \langle \widehat{t}_{ad} \phi_k,\varphi_j \rangle_{\partial T}$. Finally, the subblocks of $\mbb{M}^t$ are computed as $\mbb{M}^t_{ab,k\ell} = \langle \nu_{ab} \phi_k,\phi_\ell \rangle_{\partial T}$, with $\nu_{ab} = {\bf t}^*_a\mG{\bf t}_b$. 

The linear form can be defined as  ${\bf F}^t = [{\bf F}^t_{1}; {\bf F}^t_{2}]$, where the components of the vector are given by 
\begin{equation}\label{eq:EIM_F}
 {\bf F}^t_{a,\ell} = \langle  ({\bf t}_a\times\refN)^*\mG \be_\partial,\phi_\ell\rangle_{\partial T \cap \partial \Omega_{\br,E}} - \langle \widehat{\bt}^*_{a} \bv_\partial,\phi_\ell\rangle_{\partial T \cap \partial \Omega_{\br,V}} - \langle \bt_b^*{\bf f}_0,\phi_\ell\rangle_{\partial T \cap \partial \Omega_{\br,\rm{rad}}},
\end{equation}
for $a = 1,\,2$.

\subsection*{Empirical interpolation}

% The result of EIM is a set of $\mcal{Q}$ spatial indices $\mathscr{I}$ where $\lbrace\phi_q(x) \rbrace_{q=1}^\mcal{Q}$ are evaluated, rendering a $\mcal{Q}\times \mcal{Q}$ system from \eqref{eq:eimEIM} whereby $\bc(\upa)$ may be recovered. We direct the reader to \cite{chaturantabut2010nonlinear} for details of the algorithm.

In order to apply the EIM to the elemental matrices in \eqref{eq:elementalHDG}, we first need to identify the non-affineine parametrized functions to be interpolated, as well as the discrete set of spatial points where the interpolants will be evaluated. In the finite element context, the natural choice is the Gaussian quadrature points $\bx_\xi$ in the discretization, needed to compute the elemental inner products for volumes and faces. In addition, the non-affine functions can be readily identified from \eqref{eq:EIM_B}-\eqref{eq:EIM_F}, and involve the components of the deformation tensor $\bm G$ for the volume bilinear forms and its effect on the normal and the tangent vectors for the face bilinear forms. Details of the non-affine functions are collected in Table \ref{tab:eim}. 

For every non-affine function $f({\bf x},\upa)$, the EIM also returns the $Q$ spatial indices, denoted as $ \mathscr{I}$, where the parameter-independent orthogonal basis $\bm\Phi$ needs to be evaluated in order to retrieve the coefficients $\bm\alpha(\upa)$ for a new $\upa$, achieved through solving the $Q\times Q$ linear system
\begin{equation*}
 f({\bf x}_\mathscr{I},\upa) = \bm\Phi({\bf x}_\mathscr{I}) \bm\alpha(\upa).
\end{equation*}
Note that each non-affine function that is interpolated may require a different amount $Q$ of orthogonal basis functions to meet the EIM accuracy requirements

% For instance, the bilinear form $\mathscr{C}$ may be approximated as
% \begin{equation*}
%  \mathscr{C}(\bm G \widehat{\be},\bm \kappa) \approx \sum_{i=1}^6\sum_{q=1}^{Q^{\mathscr{C}}_i} \alpha^{\mathscr{C}}_{qi}(\upa)\mathscr{C}_q( \widehat{\be},\bm \kappa)
% \end{equation*}

\begin{table}[h!]
\footnotesize
\begin{center}
  \renewcommand{\arraystretch}{1.2}% Spread rows out...
\begin{tabular}{cccl}
\hline
Bilinear/Linear forms & Spatial points &  Non-affine functions & $\#$ Non-affine terms \\
\hline
$\mathscr{B}$  & Volume &  $\bm G_{cd},\; 1\le c \le d \le 3$ & $\sum_{i=1}^6 Q^{\mathscr{B}}_i$\\
$\mathscr{C}$  & Face & $\gamma_{cb},\; 1\le c \le 3,\, 1\le b \le 2$& $\sum_{i=1}^6 Q^{\mathscr{C}}_i$\\
$\mathscr{E}$  & Face & $\epsilon_{cb},\; 1\le c \le 3,\, 1\le b \le 2$&$\sum_{i=1}^6 Q^{\mathscr{E}}_i$\\
$\mathscr{K}$  & Face & $\kappa_{cd},\; 1\le c,d \le 3$&$\sum_{i=1}^9 Q^{\mathscr{K}}_i$\\
$\mathscr{D}$  & Face & $\delta_{cd},\; 1\le c,d \le 3$&$\sum_{i=1}^9 Q^{\mathscr{D}}_i$\\
$\mathscr{R}$  & Face &  $[\bm G (\bt_a \times \refN)]_d,\; 1\le a \le 2,\,1\le d \le 3,$&$\sum_{i=1}^6 Q^{\mathscr{R}}_i$\\
$\mathscr{L}$  & Face & $[\bm G{\bt_a}]_{d},\; 1\le a \le 2,\,1\le d \le 3$&$\sum_{i=1}^6 Q^{\mathscr{L}}_i$\\
$\mathscr{M}$  & Face &  ${\bt}^*_a\bm G\bt_b,\; 1\le a,b \le 2$&$\sum_{i=1}^3 Q_i^M$\\
$\mathscr{F}_E$  & $E$-tangent boundary  &  $ (\bt_a \times \refN)^*\bm G\be_\partial,\; 1\le a \le 2$&$\sum_{i=1}^2 Q_i^{\mathscr{F}_E}$\\
$\mathscr{F}_V$  & $V$-tangent boundary & $ {\bt^*_a}\bm G\bv_\partial,\; 1\le a \le 2$&$\sum_{i=1}^2 Q_i^{\mathscr{F}_V}$\\
\hline
\end{tabular}
\end{center}
\caption{Non-affine functions for EIM.}\label{tab:eim}
\end{table}

Once the non-affine functions are interpolated, we can approximate both $\mathscr{A}_q^\rd$ and $\mathscr{F}_q^\rd$ through affine expansions, namely

% We then set the $\mcal{N}^\rd$-dimensional space to be $\bm W^\rd_h :=\bm W_h \times\bm W_h\times \bm M_h({\bf 0})$, $\be_h^\rd := (\bv_h,\be_h,\widehat{\be}_h) $ and $\bm \xi^\rd : = (\bm \kappa,\bm \xi,\bm\mu)$. The weak formulation in \eqref{eq:hdg_weaksystemALE} may be compactly rewritten, for $\be_h^\rd \in \bm W^\rd_h $, as
% \begin{equation*}
%  \mathscr{A}^\rd(\be^\rd_h,\bm \xi^\rd\,;(\omega,\varepsilon,\upa)) = \mathscr{F}^\rd(\bm \xi^\rd\,;(\omega,\varepsilon,\upa)),\qquad \forall \bm \xi^\rd\in\bm W^\rd_h\;.
% \end{equation*}
% Applying the EIM approximations outlined in Table \ref{tab:eim} to the bilinear and linear forms in \eqref{eq:hdg_weaksystemALE} recovers a bilinear form $\mathscr{A}^\rd_{EI}\approx \mathscr{A}^\rd $ and linear functional $\mathscr{F}^\rd_{EI}\approx \mathscr{F}^\rd $ that are affine in the parameters $(\omega,\varepsilon,\upa)$, namely
\begin{align}\label{eq:EIapproximations_2}
\begin{split}
\mathscr{A}^\rd(\be^\rd, {\bm \xi}^\rd; (\omega,\varepsilon,\upa))  \approx & \mathscr{A}(\bv,\bm \kappa)  - \sum_{i=1}^6\sum_{q=1}^{Q^{\mathscr{B}}_i} \alpha^{\mathscr{B}}_{qi}(\upa)\lb \mathscr{B}_{qi}({\be},\bm \kappa) + \mathscr{B}_{qi}({\bv},\bm \xi)\rb   \\ 
& -\omega^2 \mathscr{A}_\varepsilon({\be},\bm \xi)  -\sum_{i=1}^6\sum_{q=1}^{Q^{\mathscr{C}}_i} \alpha^{\mathscr{C}}_{qi}(\upa)\mathscr{C}_{qi}( \widehat{\be},\bm \kappa)   \\
& + \sum_{i=1}^9\sum_{q=1}^{Q^{\mathscr{K}}_i}  \alpha^{\mathscr{K}}_{qi}(\upa) \mathscr{K}_{qi}({\bv},\bm \xi)+\sum_{i=1}^9\sum_{q=1}^{Q^{\mathscr{D}}_i}  \alpha^{\mathscr{D}}_{qi}(\upa) \mathscr{D}_{qi}({\be},\bm \xi)  \\
&-\sum_{i=1}^6\sum_{q=1}^{Q^{\mathscr{E}}_i}  \alpha^{\mathscr{E}}_{qi}(\upa) \mathscr{E}_{qi}( \widehat{\be},\bm \xi) -\sum_{i=1}^6\sum_{q=1}^{Q^{\mathscr{R}}_i}  \alpha^{\mathscr{R}}_{qi}(\upa) \mathscr{R}_{qi}({\bv},\bm\mu)  \\
&-\sum_{i=1}^6\sum_{q=1}^{Q^{\mathscr{L}}_i}  \alpha^{\mathscr{L}}_{qi}(\upa) \mathscr{L}_{qi}({\be},\bm\mu) + \sum_{i=1}^3\sum_{q=1}^{Q^{\mathscr{M}}_i}  \alpha^{\mathscr{M}}_{qi}(\upa) \mathscr{M}_{qi}(\widehat{\be},\bm\mu)
\end{split}\\[1ex]
\begin{split}
\mathscr{F}^\rd({\bm \xi}^\rd;(\omega,\varepsilon,\upa)) \approx  &  \sum_{i=1}^2\sum_{q =1}^{Q^{\mathscr{F}_E}_i} \alpha^{\mathscr{F}_E}_{qi}(\upa) \mathscr{F}_{E,qi}(\bm\mu) + \sum_{i=1}^2\sum_{q =1}^{Q^{\mathscr{F}_V}_i} \alpha^{\mathscr{F}_V}_{qi}(\upa)\mathscr{F}_{V,qi}(\bm\mu) \\&+\mathscr{F}_{\rm{rad}}(\bm\mu;(\omega,\varepsilon)) , \notag 
\end{split}
\end{align} 
where the parameter-independent forms $\mathscr{B}_{qi},\mathscr{C}_{qi},\mathscr{D}_{qi},\mathscr{E}_{qi},\mathscr{K}_{qi},\mathscr{R}_{qi},\mathscr{L}_{qi},\mathscr{M}_{qi},\mathscr{F}_{E,qi},\mathscr{F}_{V,qi}$ are the orthogonal spatial functions $\bm\Phi$ returned by the EIM. The expressions \eqref{eq:EIapproximations_2} are the detailed version of the more compact form given in \eqref{eq:EIapproximations}. 

% for all ${\be}^\rd = (\bv,\be,\widehat{\be})\in {\bm W}^\rd_h$, and set the HDG system with EIM approximations as
% \begin{equation}\label{eq:HsystemEI}
%  \mathscr{A}_{EI}^\rd(\be^\rd_h,\bm \xi^\rd\,;(\omega,\varepsilon,\upa)) = \mathscr{F}_{EI}^\rd(\bm \xi^\rd\,;(\omega,\varepsilon,\upa)),\qquad \forall \bm \xi^\rd\in\bm W^\rd_h\;.
% \end{equation}

% Additionally, we define an inner product for the approximation space ${\bm W}^\rd_h$ as
% \begin{equation}\label{eq:innerp}
%  ({\be}^\rd,{\bm \xi}^\rd)_{{\bm W}^\rd} = (\bv,\bm \kappa)_{\mcal{T}_h} + (\be,\bm \xi)_{\mcal{T}_h} +\langle \widehat{\be},\bm\mu \rangle_{\partial\mcal{T}_h},
% \end{equation}
% which also defines an induced norm $\norm{{\bm \xi}^\rd}_{{\bm W}^\rd} = \sqrt{({\bm \xi}^\rd, {\bm \xi}^\rd)_{{\bm W}^\rd}}$. Even though the formulation involves all the field variables, to compute the snapshots we leverage the structure of the HDG method and solve a linear system only for $\widehat{\be}_h$. We then recover the local field variables $(\bv_h,\be_h)$ element-wise using the precomputed matrices $\mbb{Z}$.

\subsection*{Computational strategy}
The linearity and recovered affine parametric dependence of the problem allow for an efficient offline-online decomposition strategy. The offline stage -- parameter independent, computationally intensive but performed only once -- comprises: (1) the computation of snapshots; (2) the POD compression that produces the orthonormalized snapshots $\snap_n, 1 \le n \le N_{\max}$ associated with the HDG approximation space at the selected parameter values; (3) the application of EIM for each bilinear form in Table \ref{tab:eim}; and (4) the formation and storage of several parameter-independent small matrices and vectors $\mathscr{B}_{qi},\mathscr{C}_{qi},\mathscr{D}_{qi},\mathscr{E}_{qi},\mathscr{K}_{qi},\mathscr{R}_{qi},\mathscr{L}_{qi},\mathscr{M}_{qi},\mathscr{F}_{E,qi},\mathscr{F}_{V,qi}$.

The online stage evaluates, for a new tuple $(\omega,\varepsilon,\upa)$, the non-affine functions at the EIM indices $\mathscr{I}$ with a cost $\mcal{O}(\mathcal{Q}\mcal{N}_{\mathscr{I}}^\rd)$ and the EIM coefficients $\bc(\upa)$ in $\mathcal{O}(2\mathcal{Q}^{\largeCdot 2})$ operations, where $\mathcal{Q}$ and $\mathcal{Q}^{\largeCdot 2}$ are given by
\begin{align*}
 \mathcal{Q} &= \sum_{i=1}^6\lb  Q^{\mathscr{B}}_i +  Q^{\mathscr{C}}_i +Q^{\mathscr{R}}_i +  Q^{\mathscr{L}}_i\rb + \sum_{i=1}^9\lb Q^{\mathscr{D}}_i +  Q^{\mathscr{K}}_i \rb+ \sum_{i=1}^3 Q_i^\mathscr{M},\\
  \mathcal{Q}^{\largeCdot 2} &= \sum_{i=1}^6\lb  \lp Q^{\mathscr{B}}_i\rp^2 +  \lp Q^{\mathscr{C}}_i\rp^2 +\lp Q^{\mathscr{R}}_i\rp^2 +  \lp Q^{\mathscr{L}}_i\rp^2\rb + \sum_{i=1}^9\lb \lp Q^{\mathscr{D}}_i\rp^2 +  \lp Q^{\mathscr{K}}_i\rp^2 \rb+ \sum_{i=1}^3 \lp Q_i^{\mathscr{M}}\rp^2.
\end{align*}
Evaluating the EIM functions has $\mcal{N}^\rd$-dependence. Nevertheless, the dimension of the set of EIM indices $\mcal{N}^\rd_{\mathscr{I}}$ where the functions are evaluated is usually much smaller than the discrete set of Gaussian quadrature points $\bx_\xi$, thus $\mcal{N}^\rd_{\mathscr{I}}\ll \mcal{N}^\rd$. Computing the trial coefficients $\lbrace \uplambda^\rd_n (\omega,\varepsilon) \rbrace_{n=1}^N$ involves assembling the linear system in \eqref{eq:EIapproximations_2} with complexity $\mcal{O}\lp \mathcal{Q}N^2\rp$ and its solution with complexity $\mcal{O}\lp N^3\rp$, hence independent of the dimension $\mcal{N}^\rd$ of the HDG approximation space. 

Finally, the dependence on $\mcal{N}^\rd$ also appears when computing $\bv_N,\,\be_N$, with a complexity of $\mcal{O}\lp 4\mcal{N}^\rd N\rp$. In the ROM community this obstacle is avoided by never evaluating the field variables; if the quantity of interest is \textit{linear} in the field variables, the offline-online strategy enables dropping the $\mcal{N}^\rd$ dependence in the online stage, see \cite{cuong2005certified,prud2002reliable}. Unfortunately, the quantities of interest in plasmonic simulations, such as the field enhancement or the optical intensity, are in general nonlinear in the field variables, thus the complexity for the online stage becomes $\mcal{N}^\rd$-dependent. Despite this shortcoming, if the quantities of interest involve localized integrals --\eg optical intensity through a surface surrounding the scatterer, or the field enhancement within sub-wavelength volumes-- we can compute $\bv_N,\,\be_N$ only for the required discretization elements, thus greatly reducing the online cost. The approximate electromagnetic fields $\bV_N,\,\bE_N$ are obtained applying \eqref{eq:transformReference} to $\bv_N,\,\be_N$ elementwise.

In summary, the implications of the above strategy are twofold: first, if $N$ and $\mcal{Q}$ are small and the quantity of interest is localized, we shall achieve very fast output evaluation, usually several orders of magnitude faster than the HDG output; second, we may choose the HDG approximation very conservatively -- to effectively eliminate the error between the exact output and HDG  output -- by only slightly affecting the online (marginal) cost. 

\section{Deformation mapping for coaxial nanogap structure}\label{app:map}

\begin{figure}[h!]
\centering
 \includegraphics[scale = .7]{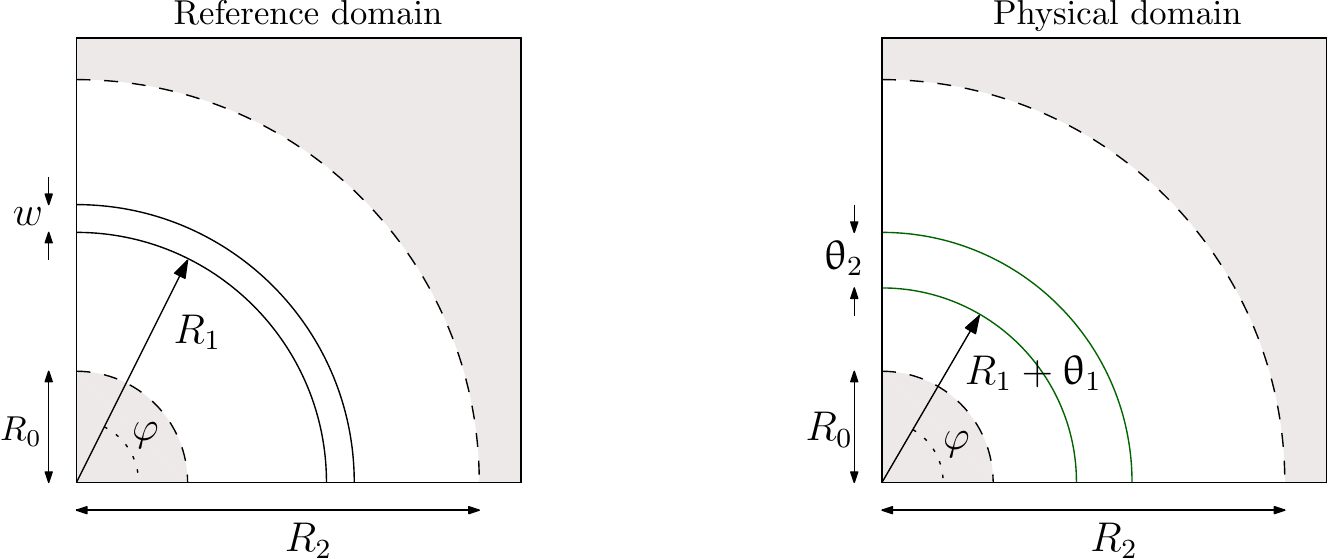}
\caption{Reference and physical domain with relevant parameters.}\label{fig:appmap}
\end{figure}

In this appendix, we introduce the deformation mapping that is employed for the periodic annular structure. For simplicity we present the mapping for only one ring of radius $R$ and gap width $w$, although the extension to multiple rings is straightforward. We therefore have two parameters $\upa = (\uptheta_1,\uptheta_2)$ that specify the modified radius $R + \uptheta_1$ and gap width $\uptheta_2$, see Fig. \ref{fig:appmap}. To derive the mapping, we resort to polar coordinates for the reference domain $(\rho_\br,\varphi_\br)$ and the physical domain $(\rho,\varphi)$. Furthermore, we have $\varphi_\br = \varphi$ since modifications occur only in the radial direction. 

The physical coordinates relate to the reference coordinates as
\begin{equation*}
 x = \mathfrak{G}_1(\refX) = \rho(\refX)\cos\varphi(\refX),\quad  y = \mathfrak{G}_2(\refX) = \rho(\refX)\sin\varphi(\refX),\quad z = \mathfrak{G}_3(\refX) = z_\br.
\end{equation*}
In order to obtain $\rho(\refX)$ we will first compute $\rho(\rho_\br,z_\br)$ and then apply the chain rule to compute the required derivatives. We propose to determine $\rho$ using $\mcal{C}^2$ cubic splines \cite{de1978practical} in the radial direction, which enable us to parametrize the curved geometry exactly without introducing additional approximations. For both reference and physical domains in Fig. \ref{fig:appmap}, the gray zones correspond to regions that remain fixed, that is $\rho = \rho_\br$, thus avoiding the singularity at the origin. 

We then use splines in the remaining regions that interpolate the knots 
\begin{equation*}
 \lbrace (\rho_{\br,i},\rho_{i}) \rbrace_{i=1}^4 = \lbrace (R_0,R_0),\,(R_1,R_1+\uptheta_1),\,(R_1+w,R_1+\uptheta_1+\uptheta_2),\,(R_2,R_2) \rbrace,
\end{equation*}
using cubic polynomials $c_i$ for each pair $(\rho_{\br,i-1},\rho_{i-1})$ and $(\rho_{\br,i},\rho_{i})$ such that $\rho = c_i(\rho_\br),\,i = 1,2,3$. To retrieve the coefficients of the cubics we impose continuity of their first and second derivatives (besides knot interpolation), rendering a spline that minimizes curvature. The extra condition to ensure a smooth blending with the gray zones is to prescribe unit slope at $R_0,\,R_2$, that is $c'_1(R_0) = c'_3(R_2) = 1$.

After solving a small 4x4 system and recovering the spline coefficients, the expression for the radial component can be compactly written as $\rho := \lbrace c_i(\rho_{\br,i}) \rbrace_{i=1}^4$. Nonetheless, this geometry modification is constant along the vertical direction, thus the mapping far from the scatterer (where radiation conditions are applied) is no longer the identity. One approach to circumvent this limitation is to use linear piecewise function $\mcal{S}$ that equals one in the region of the film and gradually vanishes as we move away from it, and express the map $\rho(\rho_\br,z_\br)$ as a convex combination of the spline and the identity, that is
\begin{equation*}
 \rho(\rho_\br,z_\br) := \mcal{S}(z_\br)\lbrace c_i(\rho_{\br,i}) \rbrace_{i=1}^4 + (1-\mcal{S}(z_\br))\rho_\br.
\end{equation*}
Finally, the required derivatives are computed invoking the chain rule, for instance
\begin{equation*}
 \dfrac{\partial x}{\partial x_\br} = \dfrac{\partial \rho}{\partial \rho_\br}\dfrac{\partial \rho_\br}{\partial x_\br}\cos\varphi + \rho \dfrac{\partial \cos\varphi}{\partial x_\br} =  \lb\mcal{S}(z_\br)\lbrace c'_i(\rho_{\br,i}) \rbrace_{i=1}^4 + 1-\mcal{S}(z_\br)\rb \dfrac{x_\br}{\rho_\br}\cos\varphi + \rho \sin\varphi\dfrac{y_r}{\rho_\br}.
\end{equation*}
It is then immediate to compute the Jacobian $\mcal{G}_{ij} = \partial_{\bx_{\br,j}} \mathfrak{G}_i(\refX)$, its determinant $g$ and the required symmetric tensor $\mG = g^{-1} \mcal{G}^{T} \mcal{G}$.

\clearpage
\newpage

\end{document}